\documentclass[prb,twocolumn,preprintnumbers,amsmath,amssymb,floatfix,longbibliography,superscriptaddress]{revtex4-1}
\usepackage{graphicx}
\usepackage{graphics}
\usepackage{psfrag}
\usepackage{hyperref}
\usepackage{multirow}
\usepackage[sort&compress]{natbib}
\usepackage{amssymb}
\usepackage{amsmath}
\usepackage{amsthm}
\usepackage{bbold}
\usepackage{bbm}
\usepackage{enumerate}
\usepackage{textcomp}

\usepackage{ulem} 

\newcommand{\vare}{\varepsilon}

\newcommand{\rmi}{{\rm i}}

\usepackage{xcolor}

\usepackage{hyperref}
\usepackage[figure,table]{hypcap}               
\hypersetup{
pdftitle = {},
pdfsubject = {},
pdfauthor = {},
pdfkeywords = {},
pdfcreator = {},
pdfproducer = {LaTeX with hyperref},
colorlinks = true,
linkcolor = blue,
anchorcolor = blue,
citecolor = blue,
filecolor = red,
menucolor = red,
pagecolor = red,
urlcolor  = blue,
breaklinks = true,
pdfstartview = FitV,
pdfhighlight = /I,
pdfpagelayout = TwoColumn,
hypertexnames=true
}

\begin{document}

\hypersetup{pdftitle={title}}
\title{d-wave superconductivity and Bogoliubov--Fermi surfaces in Rarita--Schwinger--Weyl semimetals}

\author{Julia M. Link}
\email{jmlink@sfu.ca}
\affiliation{ Department of Physics, Simon Fraser University, Burnaby, British Columbia, Canada V5A 1S6 }

\author{Igor Boettcher}
\email{iboettch@umd.edu}
\affiliation{Joint Quantum Institute, University of Maryland, College Park, MD 20742, USA}

\author{Igor F. Herbut}
\email{iherbut@sfu.ca}
\affiliation{ Department of Physics, Simon Fraser University, Burnaby, British Columbia, Canada V5A 1S6 }

\begin{abstract}
We uncover the properties of complex tensor (d-wave) superconducting order in three-dimensional Rarita--Schwinger--Weyl semimetals that host pseudospin-3/2 fermions at a fourfold linear band crossing point. Although the general theory of d-wave  order was originally developed for materials displaying quadratic band touching, it directly applies to the case of semimetals with linear dispersion, several candidate compounds of which have been discovered experimentally very recently. The spin-3/2 nature of the fermions allows for the formation of spin-2 Cooper pairs which may be described by a complex second-rank tensor order parameter. In the case of linear dispersion, for the chemical potential at the Fermi point and at strong coupling, the energetically preferred superconducting state is the uniaxial nematic state, which preserves time-reversal symmetry and provides a full (anisotropic) gap for quasiparticle excitations. In contrast, at a finite chemical potential, we find that the usual weak-coupling instability is towards the ``cyclic state", well known from the studies of multicomponent Bose-Einstein condensates, which breaks time reversal symmetry maximally, has vanishing average value of angular momentum, and features 16 small Bogoliubov--Fermi surfaces. The Rarita--Schwinger--Weyl semimetals provide therefore the first example of weakly coupled, three-dimensional, isotropic d-wave superconductors where the d-wave superconducting phase is uniquely selected by the quartic expansion of the mean-field free energy, and is not afflicted by the accidental degeneracy first noticed by Mermin over 40 years ago. We discuss the appearance and stability of the Bogoliubov-Fermi surfaces in absence of inversion symmetry in the electronic Hamiltonian, as in the case at hand.   \\[2ex]
\end{abstract}

\maketitle

The observation of exotic fermions with higher effective spin as low-energy degrees of freedom in many novel materials paves the way towards exploring quantum many-body phases at the interface of condensed matter physics and high-energy physics. Over the last years, semimetals with Fermi points that host Dirac particles \cite{GeimReview}, Weyl fermions \cite{RevModPhys.82.3045,GaoReview}, or fermions at a quadratic band touching point \cite{RevModPhys.83.1057,moon,PhysRevLett.113.106401,kondo,PhysRevLett.117.056403} have been investigated intensively. Very recent ground-breaking experiments have observed  higher-spin fermions with large topological charge and associated long Fermi arcs in surface states \cite{Bradlynaaf5037,PhysRevLett.119.206402,chang2018topological,PhysRevLett.122.076402,Rao2019ObservationOU,Sanchez,SchroeterNature,PhysRevB.99.241104,2019arXiv190412867C}. Among the candidates for an explanation of the measurements, in PdBiSe \cite{PhysRevB.99.241104} for instance, is the famous Rarita--Schwinger--Weyl (RSW) fermion with linear energy-momentum relation and the effective spin of 3/2 \cite{PhysRev.60.61,PhysRevB.93.045113,PhysRevB.94.195205,PhysRevLett.124.127602}. Although in its original Lorentz-invariant (Rarita--Schwinger (RS)) version it may be allowed to appear as an elementary particle, the RS fermion is not part of the current standard model of particle physics---although it appears in its supergravity extensions \cite{PhysRevD.13.3214}. This leaves the solid state as presently the only platform to study its fundamental properties and interactions.

The physics of spin-3/2 particles in electronic systems is potentially very rich. Here we focus on three-dimensional materials, where the time-reversal-invariant, inversion-symmetry-odd, helicity operator $h = \textbf{p}\cdot\textbf{J}$ is well defined, with momentum $\textbf{p}=(p_x,p_y,p_z)$ and $4\times 4$ spin-3/2 matrices $\textbf{J}=(J_x,J_y,J_z)$. For any low-energy $k\cdot p$--Hamiltonian $H$ constructed from $h$, helicity is a good quantum number, and so it is meaningful to associate an effective ``spin" to the corresponding fermion. The two simplest cases are comprised by the RSW Hamiltonian $H=\textbf{p}\cdot\textbf{J}$, describing a rotationally-invariant linear four-band crossing, and the celebrated Luttinger Hamiltonian $H=(\textbf{p}\cdot\textbf{J})^2-\frac{5}{4}p^2$, describing a four-fold quadratic band touching \cite{luttinger,abrikosov,abrben}. In either case, the fermion near the Fermi point is described by a four-component spinor, or, equivalently, by  four distinct fermion operators. The interaction terms that can be constructed from these could lead to several possible exotic broken symmetry  phases, including nematic order \cite{PhysRevB.92.045117,PhysRevB.93.165109,PhysRevB.95.075101,PhysRevB.97.125121}, unconventional superconductivity \cite{PhysRevB.84.220504,PhysRevB.84.220504,BayLowT,PhysRevB.93.205138,PhysRevLett.116.137001,PhysRevLett.116.177001,SmidmanReview,PhysRevLett.118.127001,PhysRevB.95.144503,PhysRevB.96.144514,PhysRevB.96.214514,PhysRevLett.120.057002,Mandal,PhysRevX.8.011029,PhysRevB.97.064504,Kimeaao4513,2018arXiv181104046S,PhysRevB.98.104514,PhysRevB.101.121301,PhysRevB.99.054505}, or  tensorial magnetism \cite{PhysRevB.85.045124,WKrempa,PhysRevX.4.041027,OngNature,PhysRevB.92.035137,PhysRevB.95.085120,PhysRevB.95.075149,BingNature,PhysRevX.8.041039}. Consequently, semimetals featuring higher-spin fermions provide a natural platform for observing novel quantum states of matter. First studies of the effects of  electron-electron interactions in RSW and RSW-like semimetals appeared in Refs.~\onlinecite{PhysRevB.93.241113,PhysRevLett.121.157602,PhysRevLett.124.127602}.

A hallmark of superconductivity in multiband systems such as those featuring spin-3/2 fermionic quasiparticles is the possibility to form Cooper pairs with an effective spin higher than unity. The superconducting order parameter, for example, can be a five-component condensate of spin 2, which is a three-dimensional spatially homogeneous ``d-wave" state. One can also think of the same d-wave order parameter as a complex second-rank irreducible tensor \cite{PhysRevLett.120.057002}. In this mathematically equivalent representation it becomes a complex symmetric traceless matrix $\phi$ which transforms as $\phi \to R \phi R^T$ under rotations $R\in\text{SO}(3)$. If $\phi$ is real, up to a possible overall phase common to all components, the corresponding superconducting state preserves  time-reversal symmetry, and obviously $\mbox{tr}( \phi^* \phi) = \mbox{tr}(\phi^2)$. In contrast, genuinely complex (again, meaning apart from an overall phase) matrices $\phi$ break time-reversal symmetry, and $\mbox{tr}( \phi^* \phi) > | \mbox{tr}(\phi^2)|$. In particular, if $\phi \neq 0$ but  $\mbox{tr}(\phi^2)=0$, we would say that the time-reversal symmetry in the d-wave superconducting state is broken {\it maximally}.

The phase diagram of complex tensor order in Luttinger semimetals with a quadratic band touching point has been investigated in Refs.~\onlinecite{PhysRevLett.120.057002,PhysRevB.97.064504, PhysRevB.100.104503, 2018arXiv181104046S}. For the chemical potential at the Fermi point ($\mu=0$ in the following), there is a phase transition at strong-coupling into a real uniaxial nematic phase. The uniaxial state, with two parallel circles of line nodes, is selected among the real order parameters, that are left degenerate by the quartic terms, by the sextic terms in the Ginzburg--Landau free energy. For $\mu>0$, on the other hand, there is  a weak-coupling second-order phase transition into a phase that breaks time-reversal symmetry. Surprisingly, just as at $\mu=0$, an expansion of the mean-field free energy to quartic order does not suffice to determine the energetically optimal ordered configuration \cite{PhysRevA.9.868}. At the level of mean-field theory, one can either employ the sextic term \cite{PhysRevLett.120.057002} in the expansion, or determine the ground state of the full mean-field BCS free energy at zero temperature exactly \cite{PhysRevB.100.104503}, with qualitatively the same, and even quantitatively very close outcomes. In particular, the latter approach yields a $C_{2\text{z}}$-symmetric superconducting ground state that breaks time-reversal symmetry almost maximally, with a high magnetization, which overlaps with the state obtained by minimization of the sextic term better than $ 99\% $. The so-called cyclic state, which breaks time reversal symmetry maximally, but in spite of that shows vanishing average magnetization, turns out to be only a local, but not the global minimum of the mean-field Ginzburg--Landau free energy at weak coupling. Thermal and quantum fluctuations have also been studied in Ref. \onlinecite{PhysRevB.97.064504}. In particular it was shown that their effect is to {\it generate} the quartic term missing in the weak-coupling derivation of the Ginzburg-Landau free energy, and with the sign that favors the formation of the cyclic state right below the critical temperature.

In this work, we study the complex tensor (spin 2) order in superconducting RSW semimetals within the mean-field approximation, and assuming a local (momentum-independent) pairing interaction. A possibility of unconventional superconductivity also arises in other semimetals.\cite{PhysRevB.99.125417, PhysRevB.101.035120, Sirohi_2019} Despite the modified band structure, the overall phase diagram resembles that of the Luttinger semimetals, but with several important novel features. For strong coupling and $\mu=0$, the superconducting state is symmetric under time reversal and again the uniaxial nematic state is selected among the real order parameters by the sextic terms in the Ginzburg--Landau free energy. In the RSW semimetals, however, the uniaxial nematic state opens a full gap for $\mu=0$. At weak coupling and $\mu>0$, the ground state again breaks the time reversal symmetry. Most importantly, and in contrast to the case of parabolic dispersion, the expansion of the mean-field free energy to quartic order in the order parameter now suffices to fully determine the cyclic state as the optimal superconducting ground state right below the critical temperature. We compare d-wave orders in Luttinger and RSW semimetals in Table \ref{Tab1}. Experimental signatures of complex tensor order would appear, for instance, in the angular or energy dependence of the optical conductivity \cite{PhysRevB.99.125146,PhysRevB.100.075104,Mandal_2019,PhysRevB.100.165115,PhysRevResearch.2.013230}, and in magnetic properties.

\renewcommand{\arraystretch}{1.6}
\begin{table}[t]
\begin{tabular}{|c|c|c|}
\hline  & Luttinger semimetals & RSW semimetals \\
 \hline \parbox{1.4cm}{band\\ structure} & \parbox{3.3cm}{four-fold quadratic\\ band touching\\ $H=(\textbf{p}\cdot\textbf{J})^2-\frac{5}{4}p^2$} & \parbox{3.3cm}{$\mbox{}$\\[1ex] four-fold linear\\ band crossing\\ $H=\textbf{p} \cdot \textbf{J}$ \\[2ex]} \\
\hline \parbox{1.4cm}{  strong coupling, \\ $\mu=0$ }  & \parbox{3.3cm}{superconducting\\ uniaxial nematic state\\ with line nodes} & \parbox{3.3cm}{$\mbox{}$\\[1ex] superconducting\\ uniaxial nematic state\\ with full gap \\[2ex]}\\
\hline \parbox{1.4cm}{ weak coupling,\\ $\mu>0$ }  &  \parbox{3.3cm}{$C_{2\text{z}}$-symmetric state\\ with broken time-\\reversal symmetry\\and BF surfaces} &  \parbox{3.3cm}{ $\mbox{}$\\[1ex] Tetragonal-symmetric\\cyclic state\\ with broken time-\\reversal symmetry\\ and BF surfaces \\[2ex]} \\
\hline
\end{tabular}
\caption{Complex tensor (d-wave) orders resulting from the mean-field theory in Luttinger and Rarita-Schwinger-Weyl semimetals.}
\label{Tab1}
\end{table}
\renewcommand{\arraystretch}{1}

Another distinguishing feature of multiband superconductors is the potential formation of Bogoliubov--Fermi (BF) surfaces in time-reversal-symmetry-breaking superconducting states, i.e. codimension-1 surfaces of gapless quasiparticle excitations \cite{PhysRevLett.118.127001,PhysRevB.96.094526,PhysRevB.96.155105,PhysRevB.98.224509,menke2019bogoliubov, Setty2020}. Potential experimental signatures of BF surfaces have recently been discussed in Ref.~\onlinecite{lapp2019experimental}. As first laid out in Ref.~\onlinecite{PhysRevLett.118.127001}, such surfaces are topologically protected in inversion-symmetric systems. They may, however, be inherently unstable towards spontaneous breaking of inversion symmetry, in presence of a favorable interactions \cite{oh2019instability}. Clearly, the single-particle Hamiltonian in RSW semimetals breaks inversion symmetry, which makes it immune to the afore-mentioned instability, but at the same time the appearance and the concomitant stability of the BF surfaces we find in all time-reversal-breaking superconducting states we checked might not be generic. Nevertheless, we report that the superconducting cyclic state which arises in RSW weakly-coupled semimetals features 16 small BF surfaces, which are non-degenerate at each value of the momentum, and hence are automatically stable in presence of weak perturbations. The properties of superconductors breaking both inversion and time-reversal symmetry recently came into focus due to applications in two-dimensional structures such as monolayer transition metal dichalcogenides or thin films of FeSe \cite{PhysRevLett.92.097001,PhysRevLett.121.157003}.

This work consists of three main sections. In Section~\ref{SecModel}, we introduce the RSW Hamiltonian, the relevant notation to study complex tensor order, and the Ginzburg-Landau theory for the d-wave superconducting state. In Section~\ref{SecStrong}, we analyse the strong coupling phase transition for $\mu=0$. In Section~\ref{SecWeak}, we eventually study the weak-coupling case, where $\mu>0$ sets the largest energy scale in the system. We then summarize our findings and give an outlook in Section~\ref{SecCon}. Some of the more technical computations are presented in the Appendices.

\section{Model}\label{SecModel}

\subsection{Rarita--Schwinger--Weyl Hamiltonian}

We consider a three-dimensional electronic system whose low-energy single-particle excitations are described by the RSW Hamiltonian
\begin{align}
 \label{mod1} H(\textbf{p}) = \textbf{p}\cdot\textbf{J},
\end{align}
with the $4\times 4$ matrices $J_i$ forming the spin-3/2 representation of the $SO(3)$, $[J_i,J_j]=\rmi \vare_{ijk}J_k$. Here and in the following we implicitly sum over repeated indices and denote $i=1,2,3=x,y,z$. In their standard representation, the spin matrices are given by
\begin{align}
 \label{mod2} J_x &= \frac{1}{2}  \begin{pmatrix} 0 & \sqrt{3} & 0 & 0 \\ \sqrt{3} & 0 & 2 & 0 \\ 0 & 2 & 0 & \sqrt{3} \\ 0 & 0 & \sqrt{3} & 0 \end{pmatrix},\\
 \label{mod3} J_y &=  \frac{\rmi}{2} \begin{pmatrix} 0 & -\sqrt{3}  & 0 & 0 \\ \sqrt{3} & 0 & -2 &0 \\ 0 & 2 & 0 & - \sqrt{3} \\ 0 & 0 & \sqrt{3} & 0\end{pmatrix},\\
 \label{mod4} J_z &= \frac{1}{2} \begin{pmatrix} 3 & 0 & 0 & 0 \\ 0 & 1 & 0 & 0 \\ 0 & 0 & -1 & 0 \\ 0 & 0 & 0 & -3 \end{pmatrix}.
\end{align}
The Hamiltonian in Eq.~\eqref{mod1} is invariant under the $\text{SO}(3)$ group of rotations. To determine its eigenvalues we may choose a frame such that $\textbf{p}=(0,0,p)$, which yields four bands with energies $+\frac{3}{2}p,\ +\frac{1}{2}p,\ -\frac{1}{2}p,\ -\frac{3}{2}p$. As shown, for instance, in Ref.~\onlinecite{PhysRevLett.124.127602}, these bands carry Chern numbers $+3,+1,-1,-3$, respectively, so that the total monopole charge of the four-fold crossing point, defined as the sum of Chern numbers of the positive energy bands, is four.

Note that in contrast to, say, Dirac particles in graphene or Weyl fermions in Weyl semimetals, a single RSW band crossing point need not imply a second RSW crossing point in the Brillouin zone. There merely need to be other topological band crossings (possibly of different type) which ensure that the total topological charge of the material is zero. For this reason we only assume the presence of a single four-fold band crossing in the following and neglect possible intervalley or nesting effects on the many-body state.

Assuming full rotation invariance as in Eq.~\eqref{mod1} is typically too strong of a constraint for modeling realistic materials, which are only invariant under discrete point groups. However, the isotropic model provides a computationally advantageous approximation for systems with small cubic anisotropy, and is also known to be a good approximation when the effects of long-range interactions  on the band structure are taken into account \cite{PhysRevB.93.241113, PhysRevB.95.075149}. Furthermore, as we show in this work, the preferred superconducting ground states that appear at both weak and strong coupling fall into irreducible representations of the cubic group, and as such are feasible orders even for anisotropic systems. Our findings should therefore be meaningful beyond the rotation-invariant limit.

In order to quantify the deviation from full rotation invariance and to further elucidate the nature of the RSW band crossing point, notice that the most general $4\times 4$ Hamiltonian linear in $\textbf{p}$ that is invariant under the cubic rotational group $O$ is given by
\begin{align}
 \label{mod5} \bar{H}(\textbf{p}) = \sum_{i=1}^3 p_i\Bigl( J_i + c J_i^3\Bigr),
\end{align}
where $c$ is a real parameter \footnote{We have $c=\frac{4(2-\alpha)}{13\alpha-14}$ for $\alpha$ defined in Eq.~(2) of Ref.~\onlinecite{PhysRevLett.124.127602}.}. This generalized version of $\bar{H}(\textbf{p})$ has been shown to emerge in critical antiperovskite Dirac materials \cite{PhysRevB.90.081112,PhysRevB.93.241113}, for various space group symmetries in three-dimensional materials \cite{Bradlynaaf5037,2019arXiv190412867C}, or in transition metal silicides \cite{PhysRevLett.119.206402}. For $c\in[-4/9,-4)$, the band structure topology is different from $H(\textbf{p})$ in Eq.~\eqref{mod1}, with the total monopole charge being $-2$. Furthermore, for $c=-4/7$ the single-particle sector acquires an enlarged Lorentz symmetry \cite{PhysRevB.93.241113} and corresponds to two copies of Weyl Hamiltonians with equal chirality \cite{Bradlynaaf5037,PhysRevLett.124.127602}. This distinguishes this limit from a massless Dirac Hamiltonian, which decomposes into two Weyl Hamiltonians with opposite chirality. The interplay between anisotropy, topology and interactions leads to intriguing effects in RSW semimetals \cite{PhysRevLett.124.127602}, but these are beyond the scope of the present study. Hereafter we consider only the fully rotation-invariant system.

\subsection{Complex tensor order}

The low-energy physics of interacting RSW fermions is assumed to be captured by the minimal Lagrangian,
\begin{align}
 \label{eq8b} L= \psi^\dagger (\partial_\tau + H(\textbf{p}) -\mu)\psi + g_1 (\psi^\dagger\psi)^2 + g_2 (\psi^\dagger\gamma_a\psi)^2,
\end{align}
which features rotation and time-reversal symmetry. Here $\psi=(\psi_{3/2}, \psi_{1/2}, \psi_{-1/2}, \psi_{-3/2})^{\rm T} $ is a four-component Grassmann field, $\tau$ denotes imaginary time, $\textbf{p}=-\rmi \nabla$ is the momentum operator, five $4\times 4$ matrices $\gamma_a$ satisfy the anticommuting Clifford algebra, $\{\gamma_a,\gamma_b\}=2\delta_{ab}$, and summation over $a=1,\dots,5$ is implied. The Fermi velocity is set to unity. (The explicit expression of the $\gamma$-matrices used here can be found in the App.~\ref{AppDiracMatrices}.) We can choose $\gamma_{1,2,3}$ to be real and $\gamma_{4,5}$ to be complex, so that the time reversal operator is given by $\mathcal{T}=\gamma_{45}\mathcal{K}$, where $\gamma_{ab}=\rmi \gamma_a\gamma_b$ and $\mathcal{K}$ denotes complex conjugation. We assume that the RSW Hamiltonian captures the band structure of an underlying material for momenta below an ultraviolet cutoff $\Lambda$, which marks the scale at which the quadratic terms in the single-particle Hamiltonian begin to matter.

In the presence of rotation symmetry, the two four-fermion interaction terms in Eq. (\ref{eq8b}) constitute a Fierz-complete set of short-range interactions \cite{PhysRevLett.113.106401,PhysRevB.92.045117, PhysRevD.100.116015}. Any other local interaction terms necessarily contain derivatives and thus are suppressed for small $\mu$. We neglect the long-range Coulomb interaction here, the subtle effects of which \cite{abrikosov, moon, PhysRevLett.113.106401,PhysRevB.92.045117} may be, due to a sufficiently large dielectric constant, assumed to set in only at the length scales much longer than the superconducting coherence length. We can rewrite the interaction terms in Eq. (\ref{eq8b}) using the Fierz rearrangement formula as $L_{\rm s}+L_{\rm d}$ with  \cite{PhysRevB.93.205138,PhysRevD.100.116015}
\begin{align}
 \label{eq9} L_{\rm s} &=  g_{\rm s} (\psi^\dagger\gamma_{45}\psi^*)(\psi^{\rm T}\gamma_{45}\psi),\\
 \label{eq10} L_{\rm d} &= g_{\rm d} (\psi^\dagger \gamma_a\gamma_{45}\psi^*)(\psi^{\rm T}\gamma_{45}\gamma_a\psi).
\end{align}
The condensation of $\Delta_{\rm s}=\langle \psi^{\rm T}\gamma_{45}\psi\rangle$ or $\Delta_a=\langle\psi^{\rm T} \gamma_{45}\gamma_a\psi\rangle$ corresponds to the onset of s- or d-wave superconductivity, respectively, and these two represent the complete set of local (momentum-independent) superconducting orders\cite{PhysRevB.93.205138}. Explicit forms of $\Delta_a$ in terms of electronic operators are given in Ref. \onlinecite{PhysRevLett.120.057002}.
We have \cite{PhysRevB.93.205138,PhysRevD.100.116015}
\begin{align}
 \label{eq11} g_{\rm s} &= \frac{1}{4}(g_1+5g_2),\\
 \label{eq12} g_{\rm d} &= \frac{1}{4}(g_1-3g_2),
\end{align}
and so an attraction in the d-wave channel can be induced by a sufficiently positive value of  $g_2$. In the following we assume $g_{\rm s}=0$ and $g_{\rm d}=-g<0$. We refer to Refs. \onlinecite{2018arXiv181104046S,PhysRevB.100.104503} for the interplay of s-wave superconductivity with d-wave order in Luttinger semimetals. The five complex components of the object $\vec{\Delta}=( \Delta_1,...,\Delta_5 )$, directly by their definition, transform the same way as the five matrices $\gamma_a$ under rotations. This is facilitated by the observation that the time-reversal operator $\mathcal{T} = \gamma_{45} \mathcal{K}$ commutes with rotations. $\gamma_a$, on the other hand, transform equivalently to the spherical harmonics of angular momentum of two \cite{abrikosov, moon, PhysRevLett.113.106401,PhysRevB.92.045117}. We may therefore collect the five $\Delta_a$ into a matrix $\phi$, which is an irreducible second-rank tensor under rotations, defined as
\begin{align}
 \label{eq5}  \phi_{ij} = \Delta_a M^a_{ij}.
\end{align}
The five real Gell-Mann matrices \cite{PhysRevB.92.045117} $M^a$ provide a basis of three-dimensional symmetric real traceless matrices. We choose the particular representation
\begin{align}
 \label{eq5b}  \phi =\begin{pmatrix} \Delta_1-\frac{1}{\sqrt{3}}\Delta_2 & \Delta_5 & \Delta_3 \\ \Delta_5 & -\Delta_1-\frac{1}{\sqrt{3}} \Delta_2 & \Delta_4 \\ \Delta_3 & \Delta_4 & \frac{2}{\sqrt{3}}\Delta_2 \end{pmatrix}.
\end{align}

The GL free energy $F(\phi)$ for complex tensor order, invariant under $\text{SO}(3)\times\text{U}(1)$, is constrained by the following remarkable fact from invariant theory: Any expansion of $F(\phi)$ in powers of $\phi$ that is invariant under $\phi \to R \phi R^T$ can only depend on the {\it eight} invariants
\begin{align}
 \nonumber I_1 &=\mbox{tr}(\phi^\dagger\phi),\ I_2 = \mbox{tr}(\phi^2),\ I_3=\mbox{tr}(\phi^\dagger{}^2),\\
 \nonumber I_4 &=\mbox{tr}(\phi^3),\ I_5 = \mbox{tr}(\phi^\dagger{}^3),\ I_6=\mbox{tr}(\phi^2\phi^\dagger),\\
 \label{eq1} I_7&=\mbox{tr}(\phi^\dagger{}^2\phi),\ I_8 = \mbox{tr}(\phi^\dagger\phi\phi^\dagger\phi),
\end{align}
which are the integrity basis of $\text{SO}(3)$. Imposing also $\text{U}(1)$-symmetry, the terms $\mathcal{O}_n$ that can appear to $n$th order in $\phi$, with $n\leq 6$, are
\begin{align}
\mathcal{O}_2 &= \{ I_1\},\\
  \mathcal{O}_4 &= \{ I_1^2,\ I_2I_3,\ I_8\},\\
\mathcal{O}_6 &= \{ I_1^3,\ I_1I_2I_3,\ I_4I_5,\ I_6I_7,\ I_1I_8\},
\end{align}
and, importantly, for $n_{\rm o}$ odd we have
\begin{align}
\mathcal{O}_{n_{\rm o}}=\varnothing.
\end{align}
These constraints result in a significant reduction of allowed terms in the Ginzburg-Landau free energy and thus a substantial computational simplification. For example, the seemingly distinct quartic term $\mbox{tr}(\phi^\dagger \phi^\dagger \phi \phi)$  is actually a linear combination of the three quartic terms in $\mathcal{O}_4$.

Of course, since $\text{SO}(3)$ has a unique irreducible complex five-dimensional representation, an equivalent way of thinking about the d-wave order parameter is as of a (pure, macroscopic) quantum state in the spin-2 Hilbert space. The five real Gell-Mann matrices then transform into each other under $\text{SO}(3)$ as the following linear combinations of the standard basis:
\begin{align}
 |M_1\rangle &= \frac{1}{\sqrt{2}}\Bigl(|-2\rangle+|2\rangle\Bigr),\\
 |M_2\rangle &=  |0\rangle,\\
 |M_3\rangle &= \frac{1}{\sqrt{2}} \Bigl(|-1\rangle-|1\rangle\Bigr),\\
 |M_4\rangle &= \frac{\rmi}{\sqrt{2}} \Bigl(|-1\rangle+|1\rangle\Bigr),\\
 |M_5\rangle &= \frac{\rmi}{\sqrt{2}} \Bigl(|-2\rangle-|2\rangle\Bigr),
\end{align}
which we label the same way as the matrices to emphasize their identical transformation law under rotations.
Here, $J_z |m\rangle = m  |m \rangle $ and $m= 0, \pm 1, \pm 2$. Note that each state $ |M_a \rangle $ is defined to be invariant under time-reversal, which is then simply complex conjugation of the coefficients of the state when expressed in this basis. One can therefore write an order parameter equivalently in this {\it real} basis as a state, defined as $|\vec{\Delta}\rangle = \Delta_a |M_a\rangle$.

We may also note that the states $|M_{1,2} \rangle $ constitute the ``$E$" two-dimensional
irreducible representation of the cubic group, whereas the remaining
$|M_{3,4,5} \rangle $ constitute the ``$T_2$" three-dimensional irreducible representation of the cubic group.

 From the above definition of the real (time-reversal-invariant) basis one readily infers that the first invariant above is
\begin{align}
I_1= 2 \Delta_a ^* \Delta_a = 2 \langle \vec{\Delta}|\vec{\Delta}\rangle,
\end{align}
i. e. simply the norm of the state, whereas the next one,
\begin{align}
I_3 = I_2^* = 2 \Delta_a ^* \Delta_a^* = 2  \langle \vec{\Delta}|\mathcal{K}| \vec{\Delta}\rangle,
\end{align}
is the overlap between the state and its time-reversed copy. A bit more algebra shows that the square of the average angular momentum, which is proportional to the magnetization, is
\begin{align}
\label{eqmagnetization} \sum_{i=1}^3 \langle\vec{\Delta}| J_i | \vec{\Delta} \rangle ^2  =  \frac{3}{2} I_8 - \frac{1}{2} I_1 ^2 - \frac{1}{4} | I_2 |^2 = \frac{1}{2} \rm tr ( [\phi, \phi^*]^2)
\end{align}
with $[ , ]$ as the commutator. The last equation demonstrates two things: 1) any real order parameter has zero average angular momentum, i.e. magnetization, 2) the three independent $\text{SO}(3) \times \text{U}(1)$ quartic terms in the Ginzburg--Landau free energy are linear combinations of the norm of the state, the $\text{SO}(3)$-invariant measure of time-reversal symmetry breaking exhibited by the state, and its average magnetization (angular momentum).

Let us now define some of the states that will play a prominent role in the mean-field Ginzburg--Landau theory that will be discussed shortly. First, if the state is real, the matrix $\phi$ can always be transformed by a rotation and a phase transformation into a diagonal form
\begin{align}
\phi = \Delta_1 M_1 +\Delta_2 M_2
\end{align}
with real $\Delta_{1,2}$. If $\Delta_1 \neq 0$, this is the biaxial nematic state, which can be at most $D_4$-symmetric, when $\Delta_2 =0$ \cite{PhysRevA.84.053616}. If, on the other hand, $\Delta_1=0$, this is the uniaxial nematic state, which is invariant under the continuous subgroup $\text{SO}(2)$ of rotations about one of the axis of the reference frame, and under rotations by $\pi$ around any orthogonal axis. As already mentioned, the average magnetization (angular momentum) of any real state is zero.

The normalized state which breaks time-reversal symmetry maximally and shows the maximal average angular momentum  of 2 is the ferromagnetic state. In the matrix notation it reads
\begin{align}
\phi =  \frac{\Delta}{\sqrt{2}} (M_1 + \rmi M_5)
\end{align}
with $\Delta$ real, or $|\vec{  \Delta }  \rangle = \Delta | 2 \rangle $  in the quantum notation. This state is also invariant under the $\text{SO}(2)$-subgroup of rotations.

Finally, in the spin-2 Hilbert space there exists a state which maximally breaks time reversal symmetry, but nevertheless has vanishing average magnetization. This is the ``cyclic state", which in the matrix notation can be written as
\begin{align}
\phi =    \frac{\Delta}{\sqrt{2}} (M_1 + \rmi M_2)
\end{align}
with $\Delta$ real. One can show that this state is invariant under the tetragonal group $T$ \cite{PhysRevA.84.053616}, which is the largest discrete subgroup of the $\text{SO}(3)$ that can be realized in the spin-2 Hilbert space. Its uniqueness, up to an $U(1)$ and $SO(3)$ transformation, follows from the observation that after the real part of the complex tensor order parameter is diagonalized, it can always be made to be proportional only to the matrix $M_1$ by a suitable $U(1)$ transformation. In order for the state to break time reversal maximally, the imaginary part of the tensor can then contain only the components $M_a$ with $a\neq 1$, and be of the same norm as the real part. The only such component that commutes with $M_1$ and by the formula in Eq. (25) would yield vanishing magnetization of the state is then $M_2$, which then implies the specific form of the cyclic state given above.

Another way to understand the uniqueness of the cyclic state is to use the Majorana representation of the spin-2 states in terms of four points on the unit sphere, and realize that, modulo an overall rotation, there is a unique way to arrange them in a tetragonally symmetric way. \cite{PhysRevA.84.053616}

\subsection{Ginzburg--Landau theory}
In this section, we present the Ginzburg--Landau expansion of the free energy near the second order phase transition towards d-wave order. In the first part we discuss how the values of the coefficients of the quartic terms in the free energy determine the superconducting state and derive the corresponding phase diagram shown in Fig. \ref{Fig1}. This phase diagram was first obtained by Mermin \cite{PhysRevA.9.868}, and we rederive it and elaborate on it here as it is needed for further presentation. In the second part we show how the coefficients in the Ginzburg--Landau free energy can be computed for the weakly coupled RSW semimetal.

\subsubsection{Phase diagram}

\begin{figure}
\centering
\includegraphics[width=8cm]{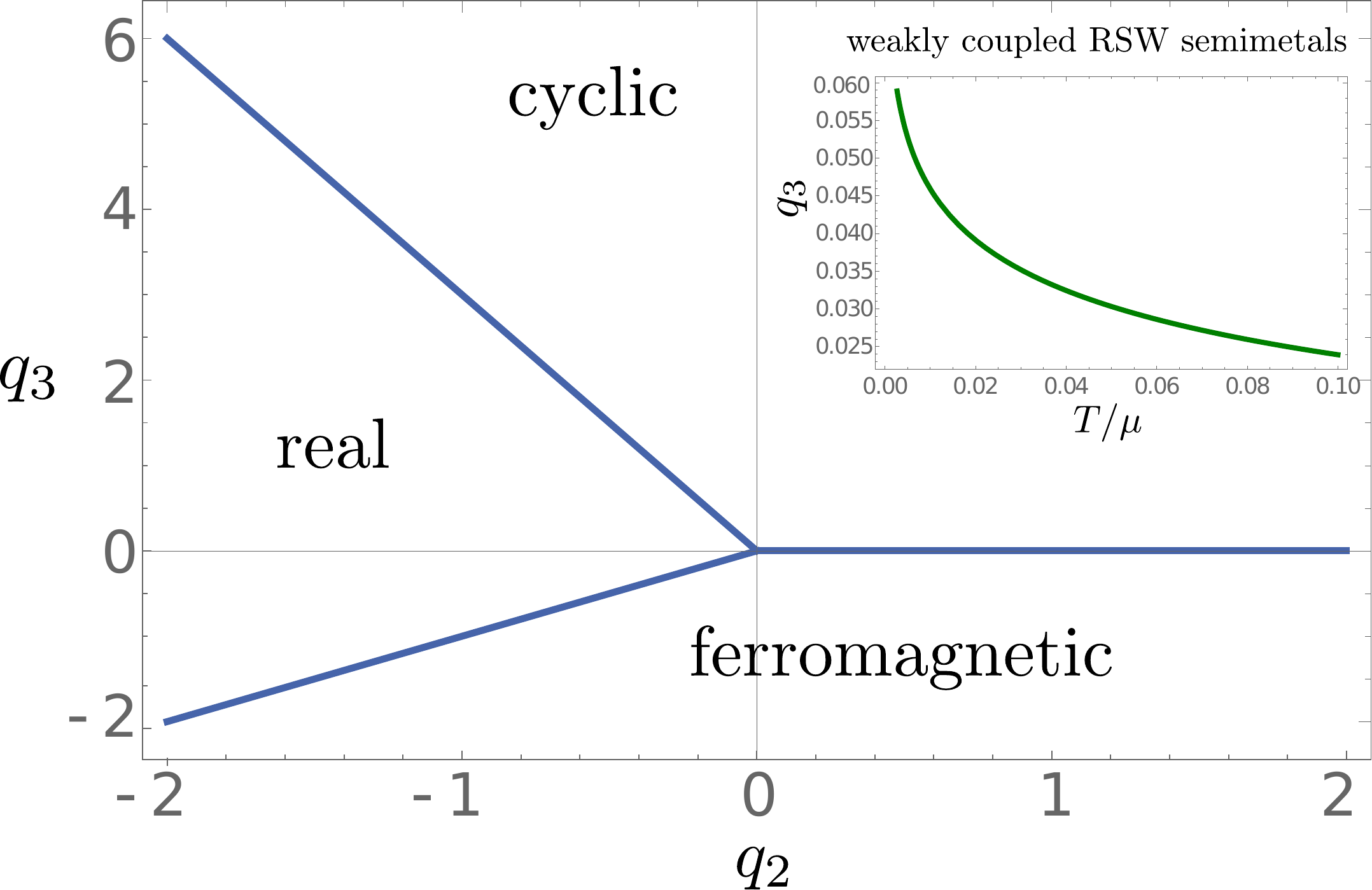}
\caption{The phase diagram of a three-dimensional d-wave superconductor is determined by the coefficients $q_2$ and $q_3$ (in arbitrary units) of the quartic terms in the Ginzburg--Landau expansion. The inset shows the $q_3$ as a function of temperature in the weak coupling regime. Positivity of both $q_2$ and
$q_3$ implies that the cyclic state is the preferred superconducting order at finite chemical potential.}
\label{Fig1}
\end{figure}

Close to a second order phase transition, we can expand the free energy $F(\vec{\Delta})$ for a uniform d-wave order parameter according to
\begin{align}
\label{mod6} F(\vec{\Delta}) = F_2(\vec{\Delta})+ F_4(\vec{\Delta}) + \mathcal{O}(\phi^6),
\end{align}
with the quadratic and  the quartic terms
\begin{align}
 \label{mod7} F_2(\vec{\Delta}) &= r |\vec{\Delta}|^2,\\
 \label{mod8} F_4(\vec{\Delta}) &= q_1 |\vec{\Delta}|^4 +q_2 |\vec{\Delta}^2|^2 + \frac{q_3}{2} I_8.
\end{align}
The ordered phase as usual corresponds to $r<0$. Crucially, although $\phi$ is a second-rank tensor under rotations, no cubic term is allowed due to the additional global $\text{U}(1)$ symmetry in the problem. Depending on the signs and relative magnitudes of $q_1$, $q_2$, and $q_3$, either real order, the cyclic state, or the ferromagnetic state yield the lowest free energy, see Fig.~\ref{Fig1}. For Luttinger semimetals, $q_3=0$ at the mean-field level, which leaves a large accidental degeneracy, broken only by the sextic terms\cite{PhysRevLett.120.057002,PhysRevA.9.868}. For RSW semimetals, however,  we will find shortly that $q_3$ is finite in general, even at the level of the one-loop approximation.

Some intuition on the interplay of the coefficients in Eq.~\eqref{mod8} can be gained by setting $q_3=0$. In order to have a stable potential bounded from below, we require $q_1>0$, and so the sign of $q_2$ determines the nature of the superconducting ground state: for $q_2<0$, a maximal value of $|\vec{\Delta}^2|$ leads to the largest decrease of the free energy in the superconducting state. This implies that $\vec{\Delta}$ is real (up to an overall phase factor) and thus preserves time-reversal symmetry. In contrast, for $q_2>0$, a minimal value of $|\vec{\Delta}^2|$ is favorable, which is readily solved by any genuinely complex order parameter satisfying $\vec{\Delta}^2=0$. Such states satisfy $\mbox{tr}(\phi^2)=0$ and break time reversal symmetry maximally, that is, time reversal transformation makes them orthogonal to themselves (Eq. (24)). As we already noted, both the cyclic and the ferromagnetic state break time reversal symmetry maximally, as well as infinitely many other states.

If $q_2 >0$ and the breaking of time reversal is preferred in the superconducting state, Eq. \eqref{eqmagnetization} shows that the sign of $q_3$ decides whether the state should or should not exhibit a finite average of the angular momentum, i. e. of magnetization.  For $q_3 <0$, maximal average magnetization is preferred and the ferromagnetic state with the maximal value of $I_8=4$ ensues. In the opposite case of $q_3 >0$, the average magnetization of the superconducting state should be minimal, although time reversal symmetry is maximally broken. The cyclic state with minimal value of  $I_8= 4/3$, which is tantamount to zero average magnetization, is then the result of these two simultaneous requirements. The phase diagram implied by the quartic terms of the Ginzburg--Landau theory is given in Fig.~\ref{Fig1}.

Importantly, for any real order parameter we have $F_4(\vec{\Delta})= (q_1+q_2+q_3)\vec{\Delta}^4$ in Eq. (\ref{mod8}). This remains true even if the coefficients $q_{1,2,3}$ are determined beyond the mean-field approximation. As a result, the expansion of the free energy to quartic order can never decide which particular real order is chosen among the many possible real states $\vec{\Delta}\in\mathbb{R}^5$. To find the optimal real state that minimizes the free energy, one would have to work with the full expression for the function $F(\vec{\Delta})$. However, expanding the latter to sextic order in the field gives a good approximation to the optimal state near critical temperature, as was shown explicitly for the  BCS d-wave state at zero temperature in Ref. \onlinecite{PhysRevB.100.104503}. The most general terms for complex tensor order to sextic order read
\begin{align}
 \nonumber F_6(\vec{\Delta}) = {}& s_1 |\vec{\Delta}|^6 + s_2 |\vec{\Delta}|^2|\vec{\Delta}^2|^2 + s_3 |\mbox{tr}(\phi^3)|^2 \\
 \label{mod9} &+ s_4 |\mbox{tr}(\phi^2\phi^\dagger)|^2 + \frac{s_5}{2} |\vec{\Delta}|^2 I_8.
\end{align}
In particular, for real order parameters we have $F_6(\vec{\Delta})=(s_1+s_2+s_5)\vec{\Delta}^6+(s_3+s_4)(\mbox{tr}\phi^3)^2$, and so the sign of $s_3+s_4$ determines whether the uniaxial nematic state, with maximal $|\mbox{tr}(\phi^3)|>0$, or the biaxial state, with minimal $|\mbox{tr}(\phi^3)|=0$ is energetically favorable.

\subsubsection{Computation of the coefficients}

The values of the coefficients entering the expansion of the free energy may be  determined in the following way.
We first compute the free energy expansion to quartic order in the field. Within the mean-field approximation we have
\begin{align}
 \label{strong1} F_2(\vec{\Delta}) &= \frac{1}{g}|\vec{\Delta}|^2 -\frac{1}{2}K_{ab}\Delta^*_a\Delta_b,\\
 \label{strong2} F_4(\vec{\Delta}) &= \frac{1}{4}K_{abcd} \Delta^*_a\Delta_b\Delta^*_c\Delta_d
\end{align}
with
\begin{align}
 \label{strong3} K_{ab} = \mbox{tr} \int_Q^\Lambda {}&G_0(\omega_n,\mu, \textbf{q})\gamma_a G_0(-\omega_n,\mu,\textbf{q})\gamma_b \:,\\
 \nonumber K_{abcd}= \mbox{tr}\int_Q^\Lambda {}&G_0(\omega_n,\mu,\textbf{q}) \gamma_a G_0(-\omega_n,\mu,\textbf{q})\gamma_b \\
 \label{strong4} &\times G_0(\omega_n,\mu,\textbf{q}) \gamma_c G_0(-\omega_n,\mu,\textbf{q})\gamma_d.
\end{align}
Here $\omega_n=(2n+1)\pi T$ denotes the fermionic Matsubara frequency with temperature $T$ and the integration comprises
\begin{align}
 \label{strong5} \int_Q^\Lambda:= T \sum_{n\in\mathbb{Z}} \int_{\textbf{q}}^\Lambda := T \sum_{n\in\mathbb{Z}} \int_{q\leq \Lambda} \frac{\mbox{d}^3q}{(2\pi)^3}
\end{align}
with ultraviolet cutoff $\Lambda\gg T,\mu$. The $4\times 4$ Gaussian propagator reads
\begin{widetext}
\begin{align}
 \nonumber G_0(\omega_n,\mu,\textbf{p}) & = \Bigl(\rmi \omega_n \mathbb{1} + H(\textbf{p})-\mu \mathbb{1}\Bigr)^{-1} \\
 \label{strong6} &= \frac{[-\rmi (\omega_n+ \rmi \mu) \mathbb{1}+H(\textbf{p})]\Bigl\{ \bigl[ (\omega_n +\rmi \mu)^2+\frac{5}{2}p^2 \bigl]\mathbb{1}-H(\textbf{p})^2\Bigr\}}{\bigl[(\omega_n+ \rmi \mu)^2+\frac{1}{4}p^2\bigr] \bigl[(\omega_n +\rmi \mu)^2+\frac{9}{4}p^2 \bigl]}
 \:.
\end{align}
\end{widetext}
In order to determine the coefficients $q_{1,2,3}$ we then insert the states $ \vec{\Delta}_1 = \Delta(0,1,0,0,0)$, $\vec{\Delta}_2 = \frac{\Delta}{\sqrt{2}}(1,\rmi,0,0,0)$, $\vec{\Delta}_3 = \frac{\Delta}{\sqrt{2}}(0,0,1,\rmi,0)$, and match them with
\begin{align}
 \label{strong10} F_4(\vec{\Delta}_1) &= (q_1+q_2+q_3)\Delta^4,
 \end{align}
 \begin{align}
 \label{strong11} F_4(\vec{\Delta}_2) &= \Bigl(q_1+\frac{2}{3}q_3\Bigr)\Delta^4,\\
 \label{strong12} F_4(\vec{\Delta}_3) &= (q_1+q_3)\Delta^4.
\end{align}
The explicit expressions for the coefficients are summarized in the App.~\ref{AppCoeff}.

\subsection{Energy spectrum of the quasiparticles}

The energy spectrum of quasiparticles has the deciding influence on the low-energy properties of superconductors. For example, line nodes in the quasiparticle excitation spectrum lead to a linear temperature dependence of the London penetration depth or the thermal conductivity. The energy spectrum of the quasiparticles is defined by the Bogoliubov--de Gennes (BdG) Hamiltonian
\begin{eqnarray}
 \label{strong15a}
 H_{\rm BdG}(\textbf{p}) &=&
 \begin{pmatrix} H(\textbf{p})-\mu \mathbb{1} & \Delta_a\gamma_a \\
 \Delta_a^*\gamma_a & - \big[ H(\textbf{p})- \mu \mathbb{1} \big]
 \end{pmatrix}
 \:,
\end{eqnarray}
which  should be understood as acting on the Nambu (eight-component) spinor $ (\psi , \mathcal{T} \psi)^{\rm T} $. Recalling that
$\mathcal{T}^{-1} H(\textbf{p}) \mathcal{T}= H^{\rm T} (-\textbf{p})$, the lower (hole) block reduces to the standard expression. In order to understand the main properties of the spectrum, let us then consider a generic case without time reversal symmetry,
\begin{equation}
\label{HBdG}  H_{\rm BdG}
 =
 \sigma_3 \otimes [H(\textbf{p}) -\mu \mathbb{1}]
 +
 \Delta_1 \sigma_1 \otimes \gamma_a + \Delta_2 \sigma_2 \otimes \gamma_b
 \:,
\end{equation}
with $\gamma_a$ and $\gamma_b$ as any of the five $\gamma$-matrices, with $a\neq b$. $\Delta_{1,2}$ are real. Since both $H(\textbf{p})$ and all $\gamma_a$ are even under time reversal, the first two terms of the Hamiltonian also respect time-reversal symmetry, \textit{i.e. } ${[\sigma_3 \otimes (H(\textbf{p})-\mu \mathbb{1}), \mathbb{1} \otimes \mathcal{T}]=[\sigma_1 \otimes \gamma_a,\mathbb{1} \otimes  \mathcal{T}]=0}$. The last term by virtue of involving the imaginary matrix $\sigma_2$, on the other hand, breaks time reversal symmetry, and
$\{\sigma_2 \otimes \gamma_b, \mathbb{1} \otimes \mathcal{T} \}=0$. Although the above $H_{\rm BdG}$ by construction breaks time reversal, one can use the antiunitary $\mathcal{T}$ to construct the operator $\mathcal{A}=\sigma_2 \otimes \mathcal{T}$, which anticommutes with all three terms in the BdG-Hamiltonian, and therefore  $\{H_{\rm BdG}, \mathcal{A} \} =0$. This, of course, only means that $H_{\rm BdG}$ has a spectrum symmetric around zero, which is the well known particle-hole symmetry inherent to Nambu's construction and any superconducting state.

Since the time reversal operator $\mathcal{T}$ inverts the momentum $\textbf{p}$ in the electron Hamiltonian $H(\textbf{p} )$, this, as well known, also means that the eigenstates of $H_{\rm BdG}$ at opposite momenta are related by the operator $\mathcal{A}$, and that the corresponding eigenvalues are similarly related as
\begin{equation}
 E(-\textbf{p}) = -E(\textbf{p})
 \:.
\end{equation}
Were the $H(\textbf{p} )$ also inversion-symmetric, the particle-hole symmetry would imply that at any fixed momentum there are pairs of eigenstates of $H_{\rm BdG}$ which differ only in sign of otherwise identical dispersion. The determinant of the Hamiltonian at any momentum, ${\det H_{\rm BdG}(\textbf{p}) =\prod_{n=1}^8 E_n(\textbf{p})}$, would then be ${\det H_{\rm BdG}(\textbf{p}) =\prod_{n=1}^4 (- E_n ^2 (\textbf{p})) }$,
and evidently non-negative. The important insight of the Ref.~\onlinecite{PhysRevLett.118.127001} was to recognize that in this situation one can nevertheless define and consider the Pfaffian, which may change sign, and thus lead to  a Bogoliubov-Fermi (BF) surface of zero energy. In the case of RSW fermions with $H(\textbf{p})=\textbf{p}\cdot \textbf{J}$, the inversion symmetry is however absent, and the particle-hole symmetry by itself does not imply that for fixed momentum $\textbf{p}$ the spectrum has to be symmetric around zero. The determinant of the Hamiltonian at a fixed momentum is therefore not automatically non-negative. We further show in the Appendix that there is actually no other operator, linear or antilinear, that would anticommute with the BdG Hamiltonian for RSW fermions if the momentum is treated as an arbitrary parameter unaffected by the transformation, when the time reversal symmetry is broken.
 The determinant therefore is actually free to change sign as a function of momentum. This permitted change of the sign of the determinant itself also allows for the appearance of a BF surface. This is what we find to be the case in explicit calculations that follow.

We may observe an additional symmetry of $H_{\rm BdG} (\textbf{p}) $ that arises at $\mu=0$. When $H(\textbf{p} )$ respects time reversal $\mathcal{T}$ but violates  inversion $\mathcal{I}$, it is odd under their combination $\mathcal{I T}$. (This is the same as saying that  $H(\textbf{p})$
becomes odd under $\mathcal{T}$ if the momentum is treated as a parameter unaffected by the transformation, as discussed above.) We may now form a new antilinear operator $\mathcal{B} = \sigma_1 \otimes \mathcal{I T} $, and observe that, since all $\gamma_a$ are even under inversion as well, the entire $H_{\rm BdG} (\textbf{p}) $ commutes with it, provided $\mu=0$. When $\mu\neq 0$ the extra term $\mu \sigma_3 \otimes \mathbb{1} $ anticommutes with $\mathcal{B}$, and the symmetry is violated. Since $\mathcal{B} ^2 = -1$, however, all the eigenstates at fixed momentum and at $\mu=0$ will be doubly degenerate, due to Kramers theorem. This will be confirmed in explicit calculations, see Fig. 3. An implication is that for $\mu=0$ the determinant $\det H_{\rm BdG}(\textbf{p}) =\prod_{n=1}^8 E_n(\textbf{p}) =(\prod_{n=1}^4 E_n(\textbf{p}))^2$, and becomes non-negative. Nevertheless, we will find that it may be still be zero on a BF surface in some, but not all, superconducting states.

\section{Strong coupling phase transition}\label{SecStrong}
We now determine the phase diagram for the chemical potential at the Fermi point ($\mu=0$). Weak short-range interactions are irrelevant due to the vanishing density of states at the band crossing point. However, for strong enough coupling, interactions induce a phase transition into a complex tensor ordered state. We show that the energetically preferred configuration is the real uniaxial nematic state with full gap.

\subsection{Phase diagram}\label{SubSecPhase}
\begin{figure}
\includegraphics[width=\columnwidth]{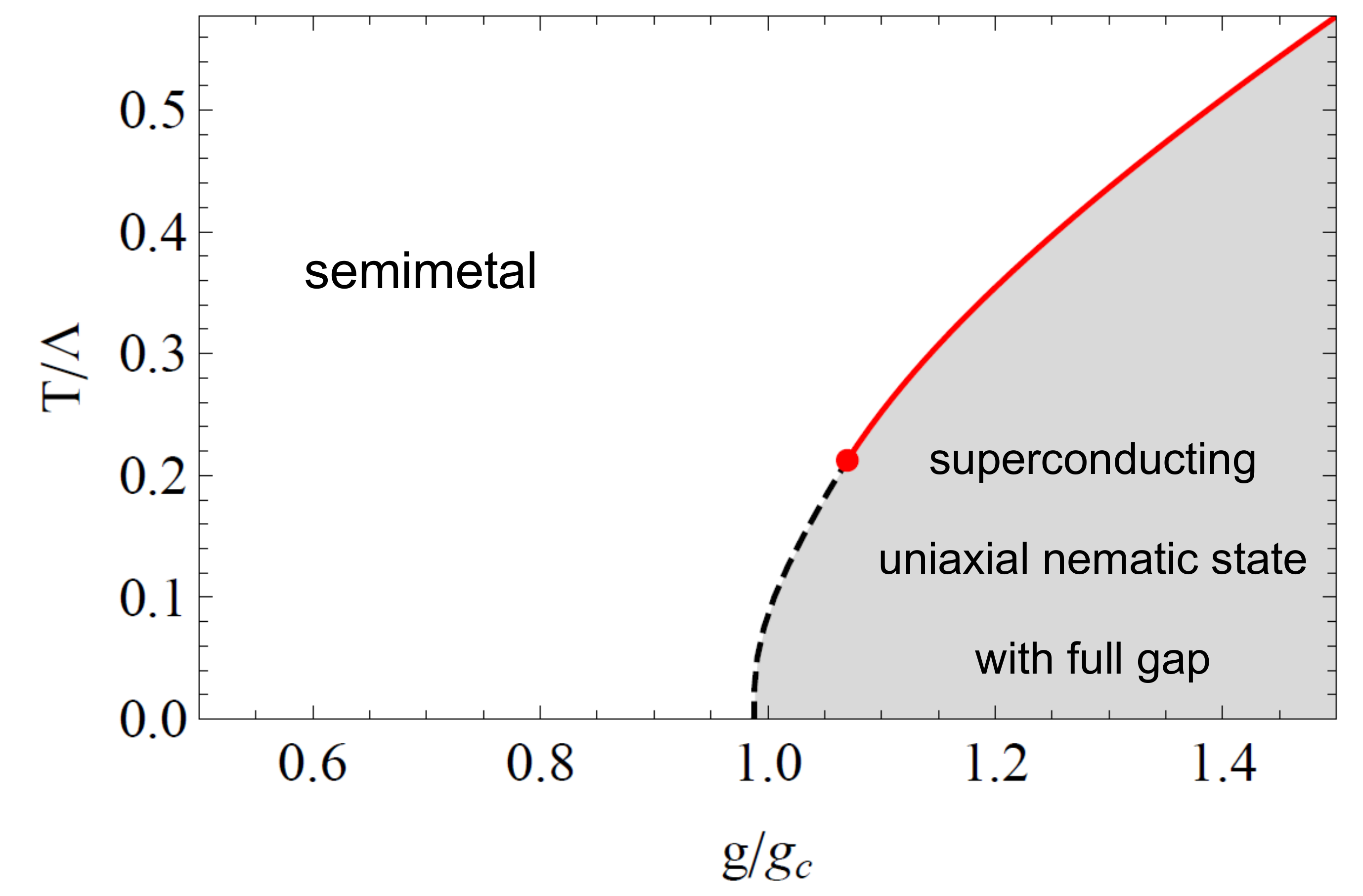}
\caption{Mean-field phase diagram for complex tensor order in RSW semimetals for $\mu=0$. We observe a strong coupling phase transition towards the superconducting state with time-reversal symmetric uniaxial nematic order parameter and full gap. The solid red and dashed black lines, respectively, correspond to a second and first order phase transition, with the tricritical point located at $(g/g_{\rm c},T_{\rm c}/\Lambda)=(1.07,0.20)$. Here $\Lambda$ is the ultraviolet cutoff of the otherwise seemingly scale-invariant system. We define $g_{\rm c}=6\pi^2/\Lambda^2$ as a reference coupling.}
\label{Fig2}
\end{figure}

In order to determine the phase diagram for $\mu=0$ we first expand the mean-field Ginzburg--Landau free energy $F(\vec{\Delta})$ to sextic order in powers of the field $\phi$. This yields a strong coupling second order phase transition at nonzero temperature. The line of second order transitions terminates at a tricritical point, where the transition turns first order and the expansion becomes meaningless, see Fig.~\ref{Fig2}.
This second order phase transition line in the plane $T/\Lambda$ vs. $g/g_{\rm c}$, with $g_{\rm c}=6\pi^2/\Lambda^2$, is determined from solving $r=r(g,T,\Lambda)=0$, where the coefficient $r$ was defined in Eqs.~\eqref{mod7} and \eqref{strong1}. At the transition point, the quartic coefficients $q_i$ in the free energy expansion are such that the symmetry-breaking order parameter is real. A stable second order transition then requires the quartic coefficient $q_1+q_2+q_3$ to be positive. From the explicit expressions we find that the second order line terminates at a tricritical point located at $(g/g_{\rm c},T_{\rm c}/\Lambda)=(1.07,0.20)$. At lower temperatures, the phase transition becomes first order, as we discuss below.

In order to determine the real order parameter that develops at the second order transition, we compute the sextic order coefficients in Eq.~\eqref{mod9} from
\begin{align}
 \label{strong13} F_6(\vec{\Delta}) &= -\frac{1}{6}K_{abcdef}\Delta_a^*\Delta_b\Delta_c^*\Delta_d\Delta_e^*\Delta_f
\end{align}
with
\begin{align}
 \nonumber K_{abcdef} ={}& \mbox{tr}\int_Q G_0(\omega_n,\textbf{q})\gamma_aG_0(-\omega_n,\textbf{q}) \gamma_b G_0(\omega_n,\textbf{q})\gamma_c \\
 \label{strong14} &\times G_0(-\omega_n,\textbf{q}) \gamma_dG_0(\omega_n,\textbf{q})\gamma_eG_0(-\omega_n,\textbf{q}) \gamma_f
\end{align}
through a matching analogous to Eqs.~\eqref{strong10}-\eqref{strong12}, see Ref.~\onlinecite{PhysRevLett.120.057002}. The resulting expressions for $s_{1,\dots,5}$ are summarized in App.~\ref{AppCoeff}. We find that $s_3+s_4$ is always negative, so that the uniaxial nematic state with maximal $\mbox{tr}(\phi^3)>0$ is energetically favored.

Below the tricritical point, the phase transition is of first order and the transition line cannot be obtained from an expansion of the mean-field free energy. Instead, for $T/\Lambda<0.20$, we employ the full expression for the mean-field free energy that follows from the quasiparticle energies for real orders. The latter are the eigenvalues of the BdG Hamiltonian at charge neutrality
\begin{equation}
 \label{strong15} H_{\rm BdG}(\textbf{p}) =
 \begin{pmatrix} H(\textbf{p}) & \Delta_a\gamma_a \\
 \Delta_a^*\gamma_a & - H(\textbf{p})
 \end{pmatrix}\:.
\end{equation}
For the most general real order $\vec{\Delta}=(\Delta_1,\Delta_2,0,0,0)$ we find doubly degenerate eigenvalues $E_{\pm}(\textbf{p}),-E_{\pm}(\textbf{p})$, with
\begin{align}
 \nonumber E_\pm(\textbf{p}) = {}& \Bigl[\frac{5}{4}p^2+\vec{\Delta}^2\pm\Bigl[p^2(p^2+2\vec{\Delta}^2)\\
 \label{strong17} &+4\Delta_1\Delta_2 d_1(\textbf{p})+2(\Delta_1^2-\Delta_2^2)d_2(\textbf{p})\Bigr]^{1/2}\Bigr]^{1/2},
\end{align}
${d_1(\textbf{p})=\sqrt{3}(p_x^2 -p_y^2)/2}$, $d_2(\textbf{p})=(2 p_z^2-p_x^2-p_y^2)/2$. We label the positive eigenvalues as $E_{1,2}=E_+$ and $E_{3,4}=E_-$. The mean-field free energy is then given by
\begin{align}
 \nonumber F(\vec{\Delta}) ={}& \frac{1}{g}|\vec{\Delta}|^2 - \frac{1}{2}\sum_{\nu=1}^4 \int_{\textbf{q}}^\Lambda\Bigl[|E_\nu(\textbf{q},\vec{\Delta})|\\
 \label{strong16} &+2T \ln (1+e^{-|E_\nu(\textbf{q},\vec{\Delta})|/T})\Bigr].
\end{align}
We verify that the uniaxial configuration ($\Delta_1=0$) has a lower free energy than the biaxial configuration ($\Delta_2=0$). The resulting phase diagram including both first and second order transitions is shown in Fig.~\ref{Fig2}.

\subsection{Nodal structure of the gap}

The nodal structure of the gap is of critical importance for low-energy transport in the superconducting state, as already noted. Furthermore, the optimal order parameter is typically such that nodes of the gap are minimized and thus a knowledge of the nodal structure of competing order parameters helps to understand the superconducting ground state.

\begin{figure*}
 \centering
  \includegraphics[width=\textwidth]{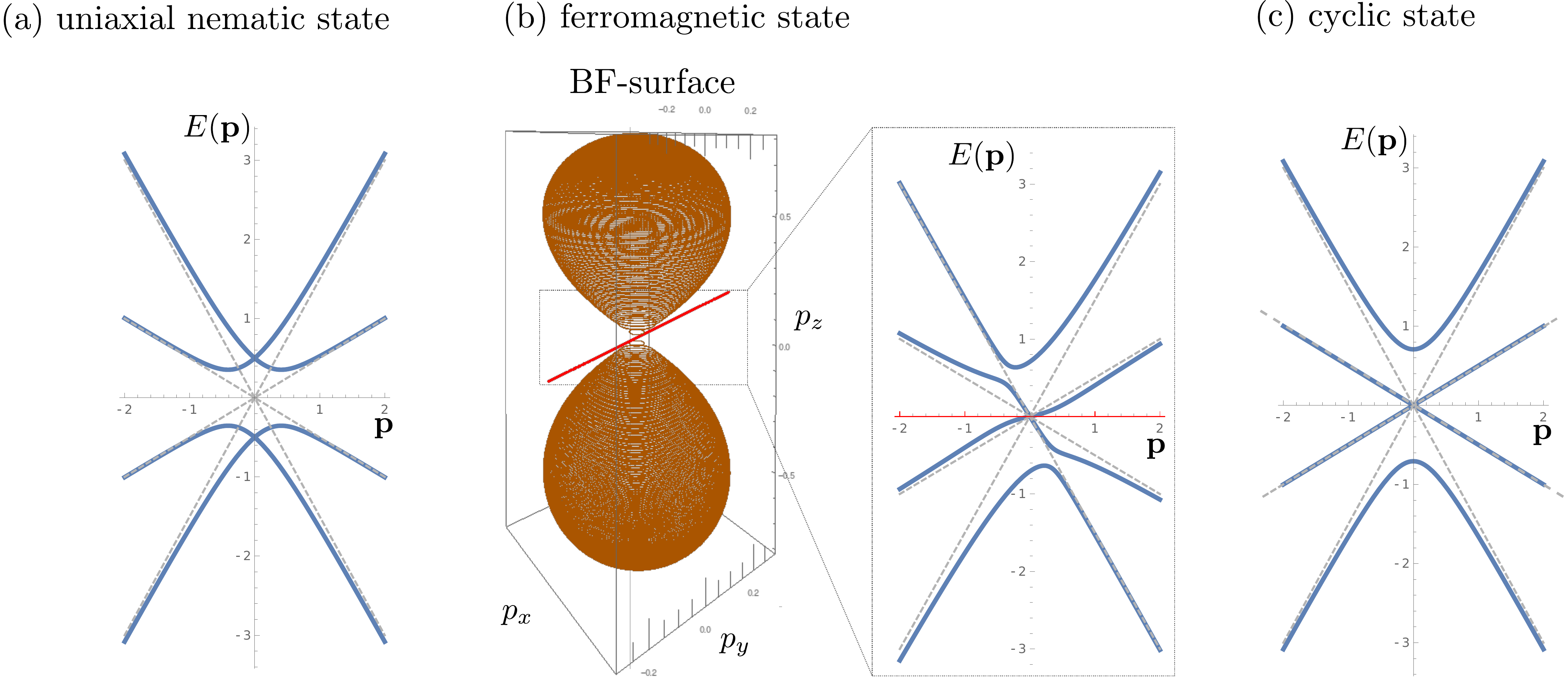}
  \onecolumngrid
  \caption{Energy dispersion of quasiparticles in competing superconducting states in the strong coupling regime for $\mu=0$. We plot the energy along the $(1,1,1)$-momentum direction for different states with $\Delta=0.5$, in units of the cutoff $\Lambda$. Figure (a) depicts the energetically preferred uniaxial nematic state whose energy spectrum is fully gapped. The superconducting states with broken time-reversal symmetry exhibit a closed dumbbell-like Bogoliubov-Fermi surface, in the ferromagnetic state, or a point node, as in the cyclic state. The grey, dashed lines in the energy spectrum denote the energy in  the normal state. The red line in panel (b) depicts the momentum direction $(1,1,1)$ along which the energy dispersion of the ferromagnetic state is plotted. The spectrum in the cyclic state plotted in (c) is actually rotationally symmetric. All the spectra at $\mu=0$ are doubly (Kramers) degenerate, as discussed in the text.}
  \label{Fig3}
 \end{figure*}

We first show that the uniaxial nematic state has a full gap, i.e. it is without nodes. For this we compute the determinant of $H_{\rm BdG}$ in Eq. (\ref{strong15}) for the uniaxial nematic state. We find
\begin{align}
 \nonumber \mbox{det}(H_{\rm BdG}(\textbf{p}))_{\rm uniaxial}={}& \frac{1}{256}\Bigl(9p^4+16\Delta^2(p^2+\Delta^2)
 \\
  \label{strong18} &+24 p^2\Delta^2\cos(2\theta_{\textbf{p}})\Bigr)^2,
\end{align}
with ${\textbf{p}=p(\sin\theta_{\textbf{p}}\cos\varphi_{\textbf{p}},\sin\theta_{\textbf{p}}\sin\varphi_{\textbf{p}},\cos\theta_{\textbf{p}})}$.
Since the determinant is the product of all quasiparticle energies in Eq.~\eqref{strong17}, a node in any of the eigenvalues at a certain momentum $\textbf{p}$ would imply a zero of the determinant. We easily verify, however, that the determinant is always strictly positive: even for $\theta$ such that $\cos(2\theta)=-1$, there is no real solution $p$ of $9p^4+16\Delta^2(p^2+\Delta^2)-24 p^2\Delta^2=0$. The resulting energy dispersion of these quasiparticles along the $(1,1,1)$ momentum direction and the gap in the energy spectrum are shown in Fig.~\ref{Fig3}(a).

From the analysis of the nodal structure of the gap we can further understand why time-reversal symmetry breaking states, such as the cyclic state or the ferromagnetic state, are disfavored for $\mu=0$. For this note that a suitable superconducting order parameter should gap out the linear band crossing at $p=0$. However, inserting $\textbf{p}=0$ into the most general form of $H_{\rm BdG}$ in Eq. (\ref{strong15}) we find
\begin{align}
 \label{strong19}  \mbox{det}(H_{\rm BdG}(\textbf{p}=0)) =|\vec{\Delta}^2|^4.
\end{align}
Consequently, states with $\vec{\Delta}^2=0$, such as, for instance, the ferromagnetic and cyclic state in Fig. 3 b) and c), do not open a gap at the crossing point and so are energetically inferior in comparison to real order parameters.

The cyclic state exhibits a point node where the $1/2$-energy bands are crossing and only the $3/2$-energy bands of the normal states are gapped out by the superconducting pairing mechanism, as shown in Fig.~\ref{Fig3}(c). Interestingly, the quasiparticle spectrum in the cyclic state happens to be also fully rotationally, and even symmetric around zero energy at fixed momentum.
For the ferromagnetic state, we find BF surfaces centered around zero momentum. The corresponding energy spectrum is not symmetric around zero energy at fixed momentum, but of course displays the general particle-hole symmetry  $E(\textbf{p})=-E(-\textbf{p})$, as can be seen in Fig.~\ref{Fig3}(b). All the spectra at $\mu = 0$ exhibit the Kramers degeneracy at any momentum, as discussed earlier.

\section{Weak coupling phase transition}\label{SecWeak}

We turn next to the phase diagram for finite chemical potential ($\Lambda\gg \mu > 0$) in the weak coupling regime. For finite chemical potential, the normal state features two spherical Fermi surfaces with radii $p_1=2\mu/3$ and $p_2=2 \mu$. Ideally, these Fermi surfaces of the normal state should be gapped out maximally in the superconducting state. We demonstrate that the uniaxial nematic state builds up line nodes at weak coupling, and that the energetically optimal configuration is the cyclic state, which breaks time-reversal symmetry and exhibits small BF surfaces.

\subsection{Phase diagram}
\begin{figure}
 \includegraphics[width=\columnwidth]{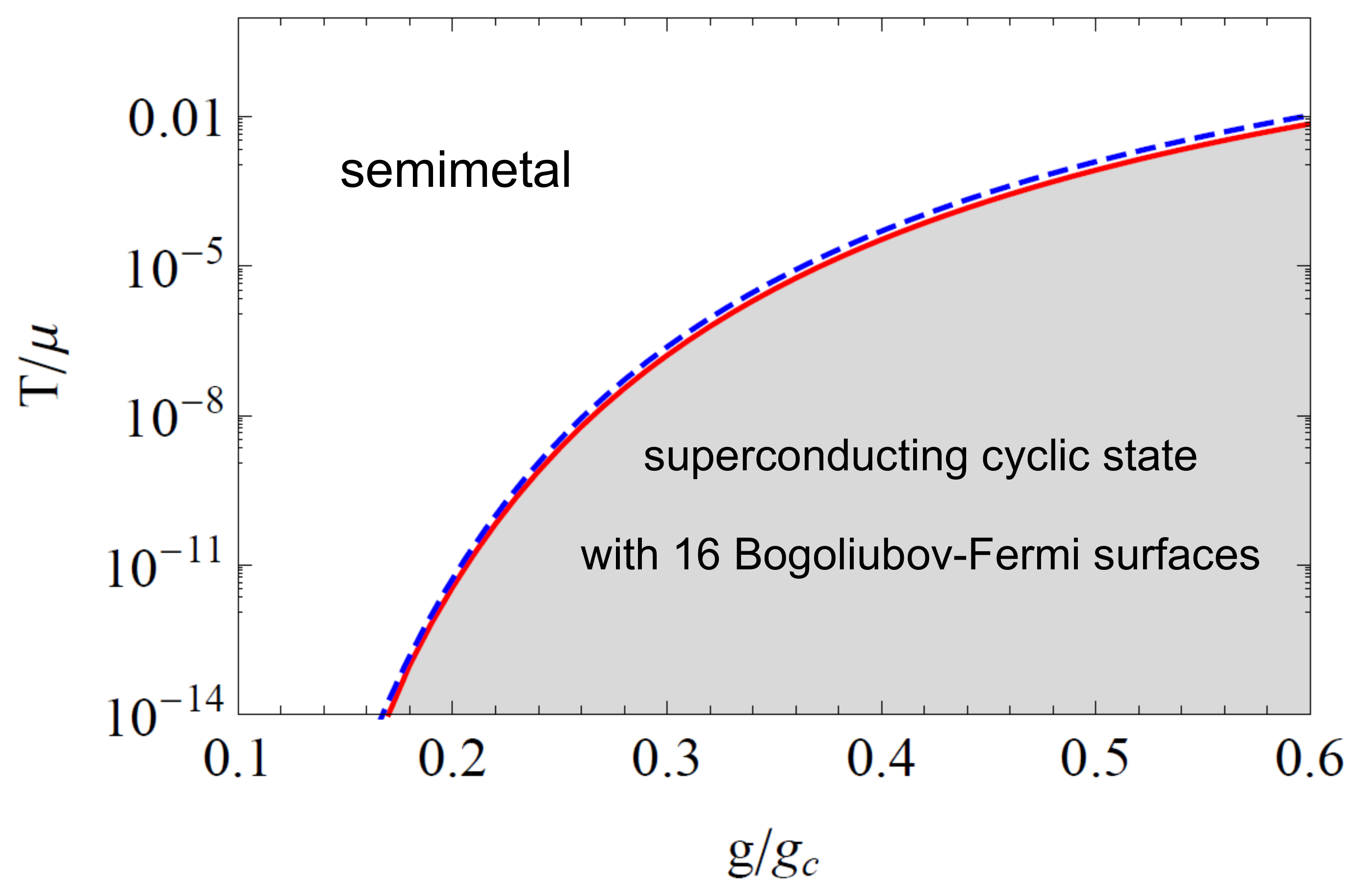}
 \caption{Phase diagram for complex tensor order in RSW semimetals with finite chemical potential $\mu > 0$. The red line shows the critical temperature for the usual BCS-like weak-coupling second order transition, the blue dashed line is the approximate formula from Eq. (\ref{eq:criticalTemp}), both extrapolated here to strong coupling. The phase transition is towards the superconducting cyclic state, which breaks time-reversal symmetry maximally, has vanishing magnetic moment, and features 16 BF surfaces. For the plot we have chosen $\Lambda/\mu=4$.}
 \label{Fig4a}
\end{figure}

Due to the nonzero value of the chemical potential $\mu$, two BCS-like instabilities occur at $p_1=2 \mu/3$ and at $p_2=2 \mu$ for arbitrarily weak attractive coupling $g>0$. This leads to a second order phase transition at the  critical temperature
\begin{align}
 \label{eq:criticalTemp} \frac{T_{\rm c}(g)}{\mu}&\simeq \exp\Bigl[-\frac{45}{112}\frac{\Lambda^2}{\mu^2}\Bigl(\frac{g_{\rm c}}{g}-1+a\frac{\mu^2}{\Lambda^2}\Bigr)\Bigr],
\end{align}
where $a=0.613$. The corresponding phase diagram for $\Lambda/\mu=4$ is shown in Fig.~\ref{Fig4a}. In order to determine the concomitant superconducting state, we compute the coefficients $q_{1,2,3}$ from mean-field theory and read off the order parameter from the phase diagram in Fig. \ref{Fig1}.

In the following, we present the leading terms in the expansions of  $q_{1,2,3}$ in the weak coupling regime, at temperatures $T\leq T_c \ll \mu$. Derivation is given  in App. \ref{App:coeff_weak}, which also contains the exact expressions in terms of frequency and momentum integrals. We find that $q_1$ is positive and has the form
\begin{equation}
 q_{1}=\frac{2^{5/3} }{1215 3^{1/6} \pi}\frac{1}{ t^{2}}+\frac{13 }{945 \pi^{2}} \ln(t) + O(t^0),
 \label{eq:q1}
\end{equation}
with $t=T/\mu$ is the small parameter near and below $T_c$. This implies that the potential is stable and bounded from below. The values of the coefficients $q_2$ and $q_3$ then select the superconducting state. These coefficients read
\begin{equation}
 q_{2}=  \frac{2^{2/3} }{1215 3^{1/6} \pi}\frac{1}{ t^{2}}+\frac{22}{189 \pi^{2}} \ln(t) + O(t^0),
 \:
 \label{eq:q2}
\end{equation}
and
\begin{align}
 q_{3}&= - \frac{31}{315 \pi^{2}} \ln(t)  + O(t^0)
 \:.
 \label{eq:q3}
\end{align}
Both coefficients are positive (as the inset in Fig.~\ref{Fig1} shows), and consequently the cyclic state that minimizes both quartic terms is favored below $T_c$. Remarkably, the coefficient $q_3$, while finite, is proportional to $\ln(1/t)$, and thus parametrically smaller than $q_1$ and $q_2$, which are both proportional to $t^{-2}$.  The RSW semimetals are a rare example of an electronic system where the d-wave superconducting state in the weak coupling regime is uniquely determined already at the quartic level of the mean field theory.

\subsection{Nodal structure of the gap}
In this section, we study the quasiparticle spectrum  for different superconducting states (uniaxial nematic, ferromagnetic, and cyclic state) in the weak coupling regime and investigate how it changes compared to the strong coupling regime.

We first show that the energetically preferred cyclic state exhibits BF surfaces. For this we compute the determinant of $H_{\rm BdG}(\textbf{p})_{\rm cyclic}$ defined in Eq.~\eqref{strong15a} with $\vec{\Delta}_{\text{cyclic}}=\frac{\Delta}{\sqrt{2}} (1,\rmi,0,0,0)$, to find
\begin{align}
 \nonumber &\det(H_{\rm BdG}(\textbf{p}))_{\rm cyclic} =  \mu^{4}\left(2\Delta^{2}+\mu^{2}\right)^{2}+\frac{81p^{8}}{256}\\
 \nonumber &+\frac{9}{16}p^{6}\left(\Delta^{2}-5\mu^{2}\right)+\frac{1}{8}p^{4} ( 2\Delta^{4}+29\Delta^{2}\mu^{2}+59\mu^{4}) \\
 & -p^{2}\left(2\Delta^{4}\mu^{2}+9\Delta^{2}\mu^{4}+5\mu^{6}\right) \\
 \nonumber &+\frac{3}{8}\Delta^{2}\mu^{2}p^{4}\Bigl[8\sin^{4}(\theta_{\textbf{p}})\cos(4\varphi_{\textbf{p}}) +4\cos(2\theta_{\textbf{p}})+7\cos(4\theta_{\textbf{p}})\Bigr].
\end{align}
 The appearance of the BF surfaces from this expression can be understood in the following way. First, it is easy to check that the function of the spherical angles featured in the square bracket in the last line reaches its minimal value (of $-31/3$) along the four diagonal directions $\textbf{p}\propto (1,1,1)$, $(-1,1,1)$, $(1,-1,1)$, and $(1,1,-1)$. Second, at $\Delta=0$ the above determinant as a function of $p$ is non-negative, with two local minima at $p_1$ and $p_2$, i. e. at the Fermi surfaces of the normal state. At these two minima the determinant vanishes. By taking a derivative with respect to $\Delta^2$ one then readily finds that the determinant must become negative in vicinity of both Fermi momenta $p_1$ and $p_2$, at and close to the above diagonal directions, for infinitesimal value of $\Delta$. These intervals of negative determinant then shrink to zero as the direction deviates sufficiently from the diagonal directions.

\begin{figure*}
\includegraphics[trim={0 5cm 0 5cm},clip,width=\textwidth]{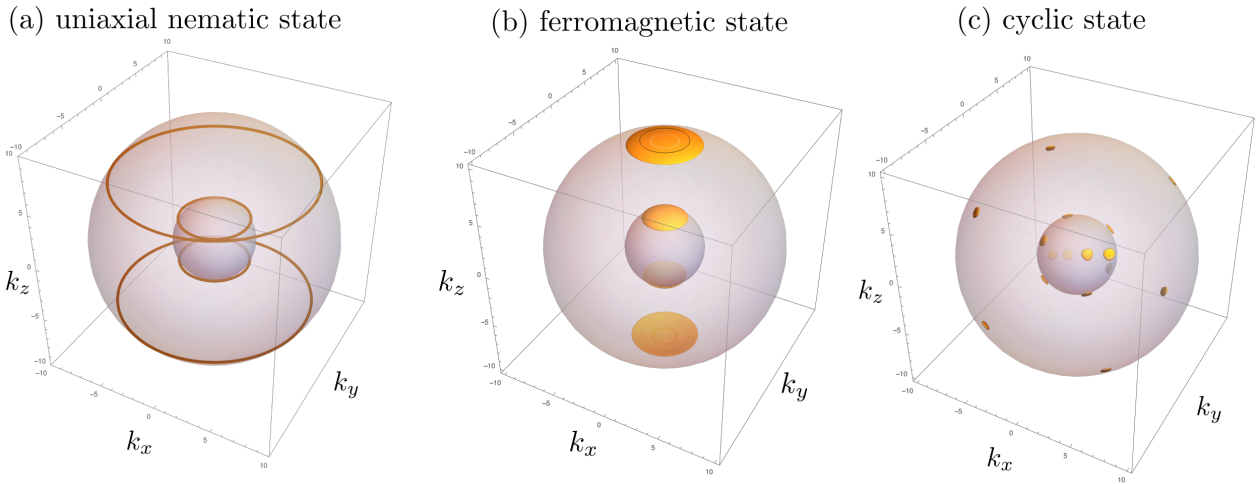}
 \caption{Nodal structure of quasiparticles in the weak coupling regime, for $\mu/\Delta=5$. The two spheres indicate the Fermi surfaces of the normal state with  radii $p_1=2\mu/3$ and $p_2=2\mu$. The real uniaxial nematic state exhibits four line nodes, which constitute two groups of parallel circles and preserve $\text{SO}(2)$ symmetry. The time-reversal-symmetry-breaking ferromagnetic state displays four thin closed surfaces along the $(0,0,1)$-direction, also featuring $\text{SO}(2)$ symmetry. The cyclic states exhibits 16 BF surfaces along the diagonals of a cube in momentum space.}
 \label{Fig4}
\end{figure*}

We therefore find (i) eight BF surfaces in the form of three-dimensional ellipsoids appearing at the first normal state Fermi sphere with $p_1=2 \mu/3$, and (ii) eight BF surfaces appearing at the second normal state Fermi sphere with $p_2=2\mu$, see Fig.~\ref{Fig4}(c).  A more detailed calculation of the allowed parameter range for the BF surfaces is presented in App.~\ref{App:cyclis_weak}. It is interesting to note that the quasiparticle spectrum appears to have the cubic symmetry, which is higher than the tetragonal symmetry of the superconducting order parameter in the cyclic state.

The other prominent state that breaks time-reversal symmetry maximally is the ferromagnetic state. An analysis of the corresponding determinant shows that the ferromagnetic state exhibits four BF surfaces aligned along to the $(0,0,1)$-direction. They are centered at the normal state Fermi surfaces of the normal state and have a bigger diameter than the BF surfaces of the cyclic state, see Fig. \ref{Fig4}(b).

In the case of the real uniaxial nematic state, the state preserves time reversal symmetry and does not exhibit BF surfaces. Instead, we find four line nodes which constitute two parallel circles on each of the two normal state Fermi surfaces, hence preserving the remaining $\text{SO}(2)$ symmetry. The line nodes occur at momenta $p_+$ and $p_-$, respectively, with an inclination $\pm\theta_{\textbf{p}}=\pm\arccos(\pm 1/\sqrt{3})$. The absolute value of the momentum $\textbf{p}$ of the four line nodes for $\Delta\ll \mu^2$ reads ${p_-=\frac{2 \mu}{3}+\frac{5 \Delta^2}{6}}$ and ${p_+=2\mu-\frac{\Delta^2}{2 \mu}}$. We display the line nodes of the uniaxial nematic state in Fig. \ref{Fig4}(a).

\section{Conclusions}\label{SecCon}

In sum, we studied the d-wave superconducting instability of the Rarita-Schwinger-Weyl (RSW) three-dimensional semimetal. A candidate material has recently been proposed in Refs. \onlinecite{PhysRevB.99.241104} and \onlinecite{schrter2019observation}. In the strong coupling regime with the chemical potential at the Fermi point $\mu=0$, the computed quartic terms of Ginzburg-Landau  free energy for the superconducting d-wave (complex tensor) state imply a time-reversal preserving (real tensor) order parameter. The next-order (sextic) terms then predict that the energetically preferred state among the real states is the uniaxial nematic, $SO(2)$--symmetric state. The uniaxial nematic state has a  fully gapped spectrum of the Bogoliubov quasiparticles.

In the weak coupling regime with finite chemical potential $\mu$, the computed quartic terms imply first that, as typically is the case, maximal breaking of time reversal is energetically preferred. The degeneracy between various time-reversal-symmetry-breaking order parameters  is now resolved already at the level of quartic terms, since we find that the term that is proportional to the state's average magnetization and which is usually missing is now small (compared to the other two terms), but positive. This implies that the d-wave state energetically preferred right below the critical temperature is the cyclic state, which breaks the time-reversal symmetry maximally, but at the same time has vanishing average magnetization. Modulo rotation and phase transformation, the cyclic state is unique such a state in the spin-2 Hilbert space of the d-wave order parameter. Its experimental manifestation may be the polar Kerr effect, showing the breaking of the time reversal symmetry, accompanied by the zero magnetization in muon spin resonance, for example.

Even though the RSW Hamiltonian is odd under inversion, the cyclic state exhibits 16 Bogoliubov-Fermi (BF) mini surfaces. We checked several other time-reversal symmetry breaking states, including the ferromagnetic state, and always found such BF surfaces to be present. We gave an argument why the absence of inversion in the electronic Hamiltonian allows for the appearance of the BF surfaces, by showing that the determinant of the BdG Hamiltonian is not necessarily positive when the time-reversal is broken in the superconducting state. We did not prove however that BF surfaces must appear in this situation, although we found no example yet of the time-reversal broken state without it. Similar result was found also in a particular time-reversal broken superconducting state in $j=1$ fermions  in Ref. \onlinecite{sim2019tripletsuperconductivity}. In contrast to BF surfaces in inversion-symmetric systems, in the systems without inversion symmetry they are here non-degenerate at a fixed momentum at the surface. This makes them immune to the further spontaneous breaking of inversion in presence of a favorable interaction recently discussed in Ref. \onlinecite{oh2019instability}. The BF surfaces should lead to modified power-laws in low-temperature magnetic field penetration depth and thermal conductivity, due to a finite, albeit rather small, density of states at zero energy.

\section{Acknowledgements.}
We gratefully acknowledge inspiring discussions with L. Janssen, T. Meng, A. Ramires, and C. Timm.
J. M. L. is supported by the DFG grant No. LI 3628/1-1.
I. B. acknowledges funding by DoE BES QIS program (award No. DESC0019449), ARO MURI, DoE ASCR Quantum Testbed Pathfinder program (award No. DESC0019040), DoE ASCR FAR-QC (award No. DE-SC0020312), NSF PFCQC program, AFOSR, ARL CDQI, and NSF PFC at JQI.
I. F. H. is supported by the NSERC of Canada.

\begin{appendix}

\section{Dirac matrices}\label{AppDiracMatrices}
We define the traceless second rank tensor
${S_{ij}=J_i J_j+J_j J_i-\frac{5}{2} \delta_{ij} \mathbb{1}_{4\times 4}}$. The five Dirac matrices $\gamma_a$ can be expressed as a combination of $S_{ij}$ and the real Gell-Mann matrices by
\begin{equation}
 \gamma_a=\frac{1}{2 \sqrt{3}} S_{ij} M_{ij}^a.
\end{equation}
The real Gell-Mann matrices are given by
\begin{align}
 \nonumber M^1 &= \begin{pmatrix} 1 & 0 & 0 \\ 0 & -1 & 0 \\ 0 & 0 & 0 \end{pmatrix},\ M^2 = \frac{1}{\sqrt{3}} \begin{pmatrix} -1 & 0 & 0 \\ 0 & -1 & 0 \\ 0 & 0 & 2 \end{pmatrix},\\
 \label{GellMann} M^3 &= \begin{pmatrix} 0 & 0 & 1 \\ 0 & 0 & 0 \\ 1 & 0 & 0 \end{pmatrix},\ M^4 = \begin{pmatrix} 0 & 0 & 0 \\ 0 & 0 & 1 \\ 0 & 1 & 0  \end{pmatrix},\ M^5 = \begin{pmatrix} 0 & 1 & 0 \\ 1 & 0 & 0 \\ 0 & 0 & 0 \end{pmatrix}.
\end{align}
For our particular choice of the matrices $J_i$, the $\gamma$-matrices are defined as
  \begin{subequations}
    \begin{align}
      \gamma_1 &= \begin{pmatrix}
                   0& 0& 1 & 0\\
                   0& 0& 0 & 1\\
                   1& 0& 0& 0\\
                   0& 1& 0 &0
                  \end{pmatrix},\
    \gamma_2 = \begin{pmatrix}
                1 & 0 & 0 & 0\\
                0& -1 &0 &0\\
                0 & 0 & -1 & 0\\
                0 & 0 & 0 & 1
               \end{pmatrix},
 \\
      \gamma_3 &= \begin{pmatrix}
                   0 & 1 & 0 & 0\\
                   1 & 0 & 0 & 0\\
                   0 & 0 & 0 & -1\\
                   0 & 0 & -1 & 0
                  \end{pmatrix},\
    \gamma_4 = \begin{pmatrix}
                0 & - \rmi & 0 & 0\\
                \rmi & 0 & 0 & 0\\
                0 & 0 & 0 & \rmi \\
                0 & 0 & - \rmi & 0
               \end{pmatrix},
  \\
      \gamma_5 &= \begin{pmatrix}
                   0 & 0 & -\rmi & 0\\
                   0 & 0 & 0 & -\rmi \\
                   \rmi & 0 & 0 & 0\\
                   0 & \rmi & 0 & 0
                  \end{pmatrix} \:.
    \end{align}
  \end{subequations}

\section{Symmetry properties of the quasiparticle excitation spectrum}\label{AppSymm}

In this section we show that for the RSW Hamiltonian $H_0(\textbf{p})=\textbf{p}\cdot \boldsymbol{J}$, without the inversion symmetry, no operator can introduce the symmetry between positive and negative energies in the superconducting state at fixed momentum, i. e. with the momentum treated simply as a parameter, if time reversal symmetry is broken, and even at $\mu=0$. In other words, we show that there is no ($\textbf{p}$-independent) operator $\mathcal{O}$, linear or antilinear, so that
\begin{equation}
\label{eqCondition} \{H_{\rm BdG}(\textbf{p}),\mathcal{O}\}=0
 \:.
\end{equation}
with $H_{\rm BdG}(\textbf{p})$ from Eq. (\ref{HBdG}), and the three numbers $\textbf{p}$ arbitrary.

(i) Let us assume the operator $\mathcal{O}$ first to be linear. Since the three $J_i$ form an irreducible representation of the $SO(3)$ algebra, Schur's lemma implies that the only operator all three generators commute with is proportional to the unit matrix. Since $J_i$ transform as a vector, the only operator that anticommutes with $J_1$ and $J_3$ is $\mathcal{U}= e^{\rmi \pi J_2}$, which then evidently commutes with $J_2$. So there is no linear operator that anticommutes with all three $J_i$. This implies that the term $\sigma_3 \otimes H_0$ anticommutes only with $\sigma_1 \otimes \mathbb{1}$ and $\sigma_2 \otimes \mathbb{1}$, which thus are the two possible candidates for $\mathcal{O}$.

The next term of the BdG Hamiltonian $\sigma_1 \otimes \gamma_a$ anticommutes with $\sigma_2 \otimes \mathbb{1}$. So if time reversal is preserved in the superconducting state, i. e. the last term in $H_{\rm BdG}$ is absent, there is a particle-hole symmetry at fixed momentum, and the determinant of the Hamiltonian is non-negative. The last, time-reversal symmetry breaking term $\sigma_2 \otimes \gamma_b$, however, commutes with
$\sigma_2 \otimes \mathbb{1}$. Hence there is no linear operator that would anticommute with the entire $H_{\rm BdG}$ at fixed momentum, in absence of time reversal symmetry, even if $\mu=0$. The same is then true at $\mu\neq 0$.

(ii) Let us next assume that the operator $\mathcal{O}$ is antilinear. For fixed momentum $\textbf{p}$, we have
\begin{equation}
 \{H_0(\textbf{p}), \mathcal{U K}\}=0,
\end{equation}
with  $\mathcal{U K}$ being the unique (time reversal) operator that anticommmutes with all three $J_i$.
Then $\sigma_3 \otimes H_0$ anticommutes with $\mathbb{1} \otimes \mathcal{U K}$ and $\sigma_3 \otimes \mathcal{U K}$:
\begin{equation}
 \{\sigma_3 \otimes H_0, \mathbb{1} \otimes \mathcal{U K}\}
 =
 \{\sigma_3 \otimes H_0, \sigma_3 \otimes \mathcal{U K}\}
 =0\:.
\end{equation}
On the other hand, all $\gamma$-matrices are even under $\mathcal{U K}$, \textit{i.e.} $[\gamma_a,\mathcal{U K}]=0$.
This implies that the second term $\sigma_1 \otimes \gamma_a$ satisfies
\begin{equation}
 [\sigma_1 \otimes \gamma_a, \mathbb{1} \otimes \mathcal{U K}]=0,\
  \{\sigma_1 \otimes \gamma_a, \sigma_3 \otimes \mathcal{U K}\}=0.
\end{equation}
However, since $\sigma_2$ is imaginary, the last term $\sigma_2 \otimes \gamma_b$ behaves precisely in the opposite way:
\begin{equation}
 \{\sigma_2 \otimes \gamma_b, \mathbb{1} \otimes \mathcal{U K}\}=0,\
 [\sigma_2 \otimes \gamma_b, \sigma_3 \otimes \mathcal{U K}]=0.
\end{equation}
So neither ${\mathbb{1} \otimes \mathcal{U K}}$ nor ${\sigma_3 \otimes \mathcal{U K}}$ anticommute with ${H_{\rm BdG}=\sigma_3 \otimes H_0 +\Delta_1 \sigma_1 \otimes \gamma_a + \Delta_2 \sigma_2 \otimes \gamma_b}$ for $\textbf{p}$ considered as fixed parameter.

We conclude that in general there is no spectral symmetry between positive and negative energies of $H_{\rm BdG}(\textbf{p})$ at a fixed generic momentum $\textbf{p}$, for systems without inversion symmetry and broken time reversal symmetry.

\section{Coefficients of the free energy expansion}\label{AppCoeff}
In this section, we give the explicit expressions for the coefficients in the Ginzburg-Landau expansion of the free energy. In the first part, we show the coefficients in the strong coupling regime with $\mu=0$. In the second part, we focus on the weak coupling regime with finite chemical potential $\mu$ and demonstrate how the analytic expression of the critical temperature and the analytic expression of the quartic coefficients $q_i$ are derived.
\begin{figure}
 \includegraphics[width=\columnwidth]{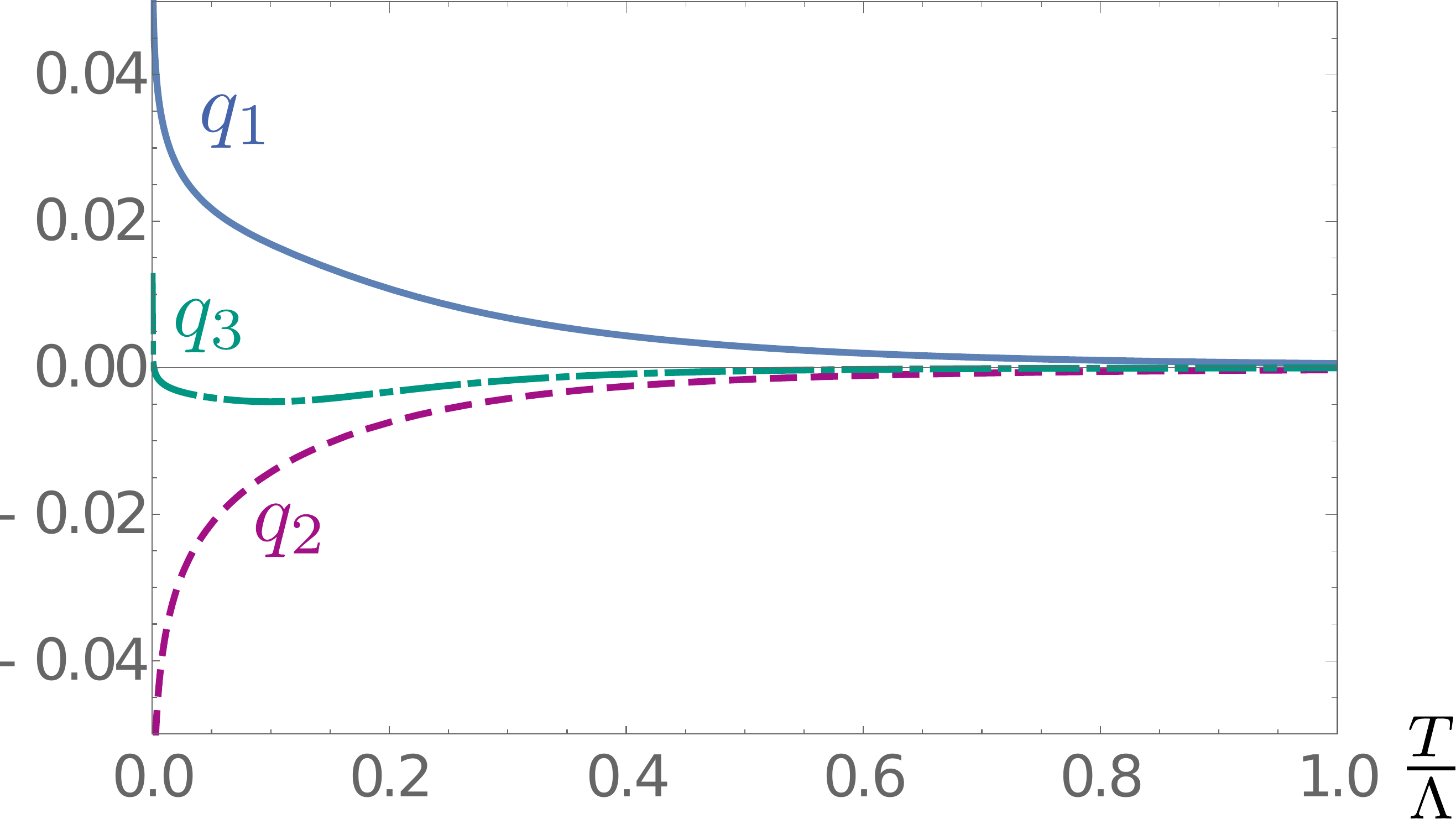}
 \caption{The quartic coefficients $q_i$ in dependence on the temperature $T$ normalized by the cut-off $\Lambda$ for $\Lambda=1$ in the strong coupling regime. The blue, solid line denotes the coefficient $q_1$, the purple, dashed the coefficient $q_2$ and the dashed-and-dotted, green line the coefficient $q_3$.}
 \label{Fig5}
\end{figure}

\subsection{Strong coupling regime}\label{AppCoeffStrong}
For $\mu=0$, the coefficient $r$ that is proportional to the quadratic expansion term of the free energy and that is defined in Eq.~\eqref{mod7}, is given by
 \begin{align}
 \nonumber  r &= \frac{1}{g} - T \sum_n \int_{\textbf{q}}^\Lambda \frac{2}{\omega_n^2+\frac{9}{4}q^2}\\
 &= \frac{1}{g} - \frac{1}{g_{\rm c}} - \Bigl(T\sum_n \int_{\textbf{q}}^\Lambda \frac{2}{\omega_n^2+\frac{9}{4}q^2} -\frac{2}{3}\int_{\textbf{q}}^\Lambda\frac{1}{q}\Bigr),
 \end{align}
where the coupling constant $g$ is positive, $g>0$ and the ``critical coupling'' $g_{\rm c}$ is defined by $g_{\rm c} =\frac{6\pi^2}{\Lambda^2}$.
The numerical values of the coefficients of the quartic terms defined in Eq.~\eqref{mod8} are determined by
 \begin{align}
  q_1 &=  T\sum_n \int_{\textbf{q}}^\Lambda \frac{2(\omega_n^4+\frac{7}{6}\omega_n^2q^2-\frac{3}{80}q^4)}{(\omega_n^2+\frac{1}{4}q^2)^2(\omega_n^2+\frac{9}{4}q^2)^2} \:,\\
 q_2 &= - T\sum_n  \int_{\textbf{q}}^\Lambda \frac{\omega_n^4+\frac{11}{6}\omega_n^2q^2+\frac{21}{80}q^4}{(\omega_n^2+\frac{1}{4}q^2)^2(\omega_n^2+\frac{9}{4}q^2)^2} \:,\\
 q_3 &= - T\sum_n  \int_{\textbf{q}}^\Lambda \frac{2q^2(\omega_n^2-\frac{3}{20}q^2)}{(\omega_n^2+\frac{1}{4}q^2)^2(\omega_n^2+\frac{9}{4}q^2)^2}.
 \end{align}
The numerical value of the coefficients $q_{1,2,3}$ in dependence on the temperature is shown in Fig.~\ref{Fig5}. We observe that $q_1$ is always positive, the coefficients $q_2$ is negative, and the value $q_3$ is such that a real ground state is selected, see Fig. \ref{Fig1}. The sextic terms in the expansion, which eventually lift the degeneracy between the real states, are given by
\begin{widetext}
\begin{align}
 s_1 &= -\frac{1}{840} T \sum_n \int_{\textbf{q}}^\Lambda \frac{2240 \omega_n^6+6160\omega_n^4q^2-3612\omega_n^2q^4+27q^6}{(\omega_n^2+\frac{1}{4}q^2)^3(\omega_n^2+\frac{9}{4}q^2)^3},\\
 s_2 &= \frac{1}{3360}  T\sum_n \int_{\textbf{q}}^\Lambda \frac{6720 \omega_n^6+9520\omega_n^4q^2-84\omega_n^2q^4+513q^6}{(\omega_n^2+\frac{1}{4}q^2)^3(\omega_n^2+\frac{9}{4}q^2)^3},\\
 s_3 &= -\frac{1}{35} T \sum_n \int_{\textbf{q}}^\Lambda \frac{(28\omega_n^2-q^2)q^4}{(\omega_n^2+\frac{1}{4}q^2)^3(\omega_n^2+\frac{9}{4}q^2)^3},\\
 s_4 &= \frac{1}{35} T \sum_n \int_{\textbf{q}}^\Lambda \frac{(28\omega_n^2-9q^2)q^4}{(\omega_n^2+\frac{1}{4}q^2)^3(\omega_n^2+\frac{9}{4}q^2)^3},\\
 s_5 &= \frac{1}{70} T\sum_n \int_{\textbf{q}}^\Lambda \frac{(560\omega_n^4-280\omega_n^2q^2-9q^4)q^2}{(\omega_n^2+\frac{1}{4}q^2)^3(\omega_n^2+\frac{9}{4}q^2)^3}
 \:.
\end{align}
Note that the integral for $s_3 +s_4$ has a negative-definite integrand. This implies that the uniaxial nematic state is energetically favored among the real order parameters.

\subsection{Weak coupling regime}\label{App:coeff_weak}

For finite chemical potential $\mu \neq 0$, the quadratic coefficient $r$ is given by
\begin{equation}
 r\left(g,\mu,T,\Lambda\right)=\frac{1}{g}-\frac{1}{2}K_{11}(T,\mu,\Lambda),
\end{equation}
where
\begin{equation}
K_{11}=T\sum_{n}\int_{\textbf{q}}^{\Lambda}\frac{8\left(128\left(\mu^{2}+\omega^{2}\right)^{3}+18q^{6}+\frac{8}{5}q^{4}\left(59\mu^{2}+95\omega^{2}\right)-32q^{2}\left(9\mu^{4}-2\mu^{2}\omega^{2}-11\omega^{4}\right)\right)}{\left[q^{2}-4(\mu-\rmi\omega)^{2}\right]\left[9q^{2}-4(\mu-\rmi\omega)^{2}\right]\left[q^{2}-4(\mu+\rmi\omega)^{2}\right]\left[9q^{2}-4(\mu+\rmi\omega)^{2}\right]}
\:.
\end{equation}
The Matsubara summation and momentum integration can be performed numerically. We can analytically approximate the integrand after Matsubara summation by expanding the integrand around the divergences at $q=\frac{2}{3}$ and $q=2$. For small $t$ we have
\begin{align}
 \nonumber K_{11} &\approx \int_{\kappa_0}^{\kappa_1}\mbox{d}q\ \text{int}_{2/3}(q) +  \int_{\kappa_1}^{\kappa_2}\mbox{d}q\ \text{int}_{2}(q) + \int_{\kappa_2}^{\Lambda}\mbox{d}q\ \text{int}_{\infty}(q)\\
 \label{K11approx} &= \frac{2}{g_{\rm c}} + \int_{\kappa_0}^{\kappa_1}\mbox{d}q\ \text{int}_{2/3}(q)+  \int_{\kappa_1}^{\kappa_2}\mbox{d}q\ \text{int}_{2}(q) - \int_0^{\kappa_2}\mbox{d}q\ \text{int}_{\infty}(q)
\end{align}
with the leading singular contributions for $t=T/\mu\to0$ to the integrand around $q\approx \frac{2}{3},\ 2,\ \infty$ given by
\begin{align}
 \text{int}_{2/3}(q) &= \frac{\mu}{\frac{135}{2}\pi^2|q/\mu-\frac{2}{3}|+90\pi^2t},\\
 \text{int}_2(q) &= \frac{\mu}{\frac{5}{2}\pi^2|q/\mu-2|+10\pi^2 t},\\
 \text{int}_{\infty}(q) &= \frac{q}{\frac{3}{2}\pi^2}
 \:.
\end{align}
We then find
\begin{align}
 \nonumber K_{11} &\approx \frac{2}{g_{\rm c}} -\frac{\kappa_2^2}{3\pi^2} + \frac{2\mu^2}{135\pi^2} \Bigl[\ln\Bigl(1+\frac{2-3\frac{\kappa_0}{\mu}}{4t}\Bigr)+\ln\Bigl(1+\frac{3\frac{\kappa_1}{\mu}-2}{4t}\Bigr)\Bigr] + \frac{2\mu^2}{5\pi^2}\Bigl[\ln\Bigl(1+\frac{2-\frac{\kappa_1}{\mu}}{4t}\Bigr)+\ln\Bigl(1+\frac{\frac{\kappa_2}{\mu}-2}{4t}\Bigr)\Bigr]\\
 \label{K11approx2} &= -\frac{112\mu^2}{135\pi^2}\ln t + \frac{2}{g_{\rm c}}- \frac{a}{3\pi^2} \mu^2
 \:,
\end{align}
where the divergent term $\propto \ln t$ is universal and independent of the approximations made to the finite part of the integral. In contrast, the $t$-independent term $a$ is non-universal. We fix $a$ by comparing the right-hand side to the numerically evaluated $K_{11}$ to be $a=0.613$. This leads to
\begin{align}
  r\left(g,\mu,T,\Lambda\right) \simeq \frac{1}{g}- \frac{1}{g_{\rm c}} +\frac{a}{6\pi^2}\mu^2 +\frac{56\mu^2}{135\pi^2} \ln t
\end{align}
and from $r(g,\mu,T_{\rm c},\Lambda)=0$ we deduce
\begin{align}
 \nonumber \frac{T_{\rm c}}{\mu} &\simeq \exp\Bigl[ - \frac{135\pi^2}{56\mu^2 g_{\rm c}}\Bigl(\frac{g_{\rm c}}{g}-1+\frac{a}{6\pi^2} g_{\rm c}\mu^2\Bigr)\Bigr]\\
 &=\exp\Bigl[-\frac{45}{112}\frac{\Lambda^2}{\mu^2}\Bigl(\frac{g_{\rm c}}{g}-1+a\frac{\mu^2}{\Lambda^2}\Bigr)\Bigr]
\end{align}
as quoted in Eq. (\ref{eq:criticalTemp}).

The quartic coefficients for finite $\mu$ are given by
\begin{align}
q_{1}(T,\mu) & =-T\sum_{n }\int_{\textbf{q}}^{\Lambda}\frac{32}{105\left[q^{2}-4(\mu-\rmi\omega)^{2}\right]^{2}\left[9q^{2}-4(\mu-\rmi\omega)^{2}\right]^{2}\left[q^{2}-4(\mu+\rmi\omega)^{2}\right]^{2}\left[9q^{2}-4(\mu+\rmi\omega)^{2}\right]^{2}}\times
\nonumber \\
\times & [-430080\left(\mu^{2}+\omega^{2}\right)^{6}+5103q^{12}-648q^{10}\left(373\mu^{2}+175\omega^{2}\right)-48q^{8}\left(4189\mu^{4}+72906\mu^{2}\omega^{2}+29757\omega^{4}\right)
\nonumber \\
 & +256q^{6}\left(6949\mu^{6}+39889\mu^{4}\omega^{2}-13141\mu^{2}\omega^{4}-18865\omega^{6}\right)
 \nonumber \\
 & -1792q^{4}\left(\mu^{2}+\omega^{2}\right)^{2}\left(1529\mu^{4}-3902\mu^{2}\omega^{2}+3161\omega^{4}\right)+71680q^{2}\left(\mu^{2}+\omega^{2}\right)^{4}\left(25\mu^{2}-37\omega^{2}\right)]\:,
 \label{eq:q1fullint}\\
q_{2}(T,\mu) & =-T\sum_{n}\int_{\textbf{q}}^{\Lambda}\frac{16}{105\left[q^{2}-4(\mu-\rmi\omega)^{2}\right]^{2}\left[9q^{2}-4(\mu-\rmi\omega)^{2}\right]^{2}\left[q^{2}-4(\mu+\rmi\omega)^{2}\right]^{2}\left[9q^{2}-4(\mu+\rmi\omega)^{2}\right]^{2}}\times
\nonumber \\
\times & [430080\left(\mu^{2}+\omega^{2}\right)^{6}+35721q^{12}-648q^{10}\left(967\mu^{2}-875\omega^{2}\right)+48q^{8}\left(30317\mu^{4}-10518\mu^{2}\omega^{2}+66381\omega^{4}\right)
\nonumber \\
 & -256q^{6}\left(7169\mu^{6}-3571\mu^{4}\omega^{2}-24401\mu^{2}\omega^{4}-29645\omega^{6}\right)
 \nonumber \\
 & +1792q^{4}\left(\mu^{2}+\omega^{2}\right)^{2}\left(1633\mu^{4}-2542\mu^{2}\omega^{2}+4033\omega^{4}\right)-71680q^{2}\left(29\mu^{2}-41\omega^{2}\right)\left(\mu^{2}+\omega^{2}\right)^{4}]\:,
 \label{eq:q2fullint}
 \\
 \label{eq:qi_strong_coupling} q_{3}(T,\mu) & =+T\sum_{n}\int_{\textbf{q}}^{\Lambda}\frac{128\:q^{2}}{35\left[q^{2}-4(\mu-\rmi\omega)^{2}\right]^{2}\left[9q^{2}-4(\mu-\rmi\omega)^{2}\right]^{2}\left[q^{2}-4(\mu+\rmi\omega)^{2}\right]^{2}\left[9q^{2}-4(\mu+\rmi\omega)^{2}\right]^{2}}\times
\nonumber \\
\times & [35840\left(\mu^{2}-\omega^{2}\right)\left(\mu^{2}+\omega^{2}\right)^{4}+1701q^{10}+108q^{8}\left(93\mu^{2}+35\omega^{2}\right)-96q^{6}\left(1277\mu^{4}+1418\mu^{2}\omega^{2}+637\omega^{4}\right) \nonumber \\
 & +128q^{4}\left(2047\mu^{6}+4327\mu^{4}\omega^{2}+857\mu^{2}\omega^{4}-1855\omega^{6}\right)-1792q^{2}\left(\mu^{2}+\omega^{2}\right)^{2}\left(97\mu^{4}+50\mu^{2}\omega^{2}+97\omega^{4}\right)]
 \:.
\end{align}
The integrals are convergent and readily evaluated numerically. In order to find an analytical expression, we first introduce dimensionless variables with $k=q/\mu$ and $t=T/\mu$. Then we perform a zero-temperature frequency integration and then expand around $k=2/3$ and $k=2$. This way, we can determine the divergence of the integrand, \textit{i.e.} whether the integrand is proportional to $1/k^3$, $1/k^2$, or $1/k$. Next, we perform the Matsubara summation for the original expressions in Eqs.~\eqref{eq:q1fullint},~\eqref{eq:q2fullint} and \eqref{eq:qi_strong_coupling} and expand them around $k=2/3$ and $k=2$, in order to extract the temperature dependence of the integrand at these singular points. With this knowledge, we can make an appropriate approximate ansatz for the integrands of $q_{1,2,3}$ and establish the analytic divergence structure of the coefficients. The integrand of $q_1$ is described by
\begin{align}
\label{eq:q1_int} \rm \text{int}_{q_{1}}(k \approx 2/3) & =\frac{1}{8505\pi^{2}|k-\frac{2}{3}|^{3}/2+15120\pi^{2}t^{3}}\Bigl[1-\frac{45}{4}\Bigl(k-\frac{2}{3}\Bigr)-\frac{117}{4}\Bigl(k-\frac{2}{3}\Bigr)^{2} \Bigr],\\
\rm \text{int}_{q_{1}}(k \approx 2) & = \frac{1}{35\pi^{2}|k-2|^{3}/2+1680\pi^{2}t^{3}}\Bigl[1+\frac{3}{4}\left(k - 2\right)\Bigr],\\
 \text{int}_{q_{1}}( k \approx \infty) &=  \frac{1}{15\pi^{2} k}.
\end{align}
For the integrand of $q_2$ we find
\begin{align}
\label{eq:q2_int}\rm int_{q_{2}}(k\approx \frac{2}{3})
&=  \frac{1}{8505\pi^{2} \left| k-\frac{2}{3} \right|^{3}+30240\pi^{2}t^{3}} \Bigl[1+9\Bigl(k-\frac{2}{3}\Bigr)-\frac{261}{2}\Bigl(k-\frac{2}{3}\Bigr)^{2}\Bigr],\\
\rm int_{q_{2}}(k \approx 2)= & \frac{1}{35\pi^{2}\left|k-2\right|^{3}+3360\pi^{2}t^{3}}\Bigl[1+\left(k-2\right)-\frac{3}{2}\left(k-2\right)^{2}\Bigr],\\
\rm int_{q_{2}}(k \approx \infty) & =-\frac{1}{10 \pi^2 k}.
\end{align}
At last, the coefficient $q_3$ is given by the functions
\begin{align}
 \label{eq:q3_int} \rm int_{q_3}(k \approx 2/3)
 &=
 \frac{-1}{1890 \pi^2 [\left(k -\frac{2}{3} \right)^2+2 t^2]} \frac{k-\frac{2}{3}}{\sqrt{\left(k-\frac{2}{3} \right)^2 + 2 t^2}}
 +
 \frac{2}{315 \pi^2 [\sqrt{\left(k-\frac{2}{3} \right)^2} + t]},\\
\rm int_{q_3}(k \approx 2)
 &=
 \frac{3}{70 \pi^2 [(k-2)^2 + 18 t^2 ]} \frac{k-2}{\sqrt{(k-2)^2+18 t^2}}
 +
 \frac{3}{70 \pi^2 [ \sqrt{(k-2)^2} +t ]} ,\\
 \rm int_{q_3}(k \approx \infty) & =\frac{1}{90\:k\pi^2}
 \:.
 \label{eq:q3_int_inf}
\end{align}
After performing the momentum integration in Eqs. (\ref{eq:q1_int}-\ref{eq:q3_int_inf}) with these approximations, and fitting the non-divergent part to match the full expression as in Eq.~\eqref{K11approx2}, we obtain the analytic expressions for the coefficients $q_{1,2,3}$ given in Eqs.~\eqref{eq:q1}, \eqref{eq:q2}, and \eqref{eq:q3}.

\section{Nodal structure of the superconducting gap}
In this Appendix, we derive the nodal structure of the superconducting gap in the weak coupling limit for $\mu\neq 0$ for the three relevant superconducting orders due to Fig. \ref{Fig1}: the real uniaxial nematic state, and the time-reversal symmetry breaking ferromagnetic and cyclic state. While the former case exhibits four circular line nodes, the latter two states feature BF surfaces.

\subsection{Uniaxial nematic state}
The determinant of the uniaxial state with ${\vec{\Delta}_{\rm uniaxial}}=\Delta(0,1,0,0,0)$ is
\begin{equation}
\det(H_{\rm BdG}(\textbf{p}))_{\rm uniaxial}=\left(\frac{9}{16}\right)^{2}\left[p^{4}+(A+B)p^{2}+\frac{16}{9}(\Delta^{2}+\mu^{2})^{2}\right]\left[p^{4}+(A-B)p^{2}+\frac{16}{9}(\Delta^{2}+\mu^{2})^{2}\right]
\end{equation}
 with
\begin{align}
A= & \frac{8}{9}(2\Delta^{2}-5\mu^{2}+3\Delta^{2}\cos(2\theta_{\textbf{p}})) \:,\\
B= & \frac{8}{9}\sqrt{2}\rmi\Delta\mu\sqrt{11+12\cos(2\theta_{\textbf{p}})+9\cos(4\theta_{\textbf{p}})}
\:.
\end{align}
Any zero $p$ of the determinant must satisfy
\begin{equation}
p^{2}=-\frac{A+B}{2}\pm\frac{1}{2}\sqrt{(A+B)^{2}-\frac{64}{9}(\Delta^{2}+\mu^{2})^{2}}
\:.
\end{equation}
In order for $p^{2}$ to be real, $B=0$, which yields $\theta_{\textbf{p}}\equiv\theta_0$ with
\begin{equation}
 11+12\cos(2\theta_0)+9\cos(4\theta_0)=0\Leftrightarrow\theta_0=\arccos\left(\pm\frac{1}{\sqrt{3}}\right)
 \:.
\end{equation}
This leads to the condition
\begin{equation}
 \label{EqpPlus} p^{2}=\frac{4}{9}\left(-\Delta^{2}+5\mu^{2}\pm2\sqrt{-2\Delta^{4}-7\Delta^{2}\mu^{2}+4\mu^{4}}\right)
 \:.
\end{equation}
 Hence we have four nodal loops for the uniaxial state located at
\begin{equation}
 (p_{\pm},\theta_{0}),\quad(p_{\pm},-\theta_{0})
\end{equation}
 for every $\phi\in[0,2\pi)$. For $\Delta\ll \mu$ we expand Eq.~ \eqref{EqpPlus}, and obtain $p_{+}=2\mu-\frac{\Delta^{2}}{2\mu}$, $p_{-}=\frac{2\mu}{3}+\frac{5\Delta^{2}}{6}$.

\subsection{Ferromagnetic state}
The determinant for the ferromagnetic state with $\vec{\Delta}_{\rm ferromagnetic}=\frac{\Delta}{\sqrt{2}}(1,0,0,0,i)$ is
\begin{align}
\det(H_{\rm BdG}(\textbf{p}))_{\rm ferromagnetic} & =\mu^{4}\left(2\Delta^{2}+\mu^{2}\right)^{2}+\frac{81p^{8}}{256}-\frac{45\mu^{2}p^{6}}{16}+\frac{1}{8}p^{4}\left(9\Delta^{4}+36\Delta^{2}\mu^{2}+59\mu^{4}\right) \nonumber \\
 & -p^{2} \left(8 \Delta^{4} \mu^{2}+10 \Delta^{2} \mu^{4}+5 \mu^{6}\right)\\
 & +\frac{1}{16}\Delta^{2}p^{2}\left(6p^{2}\left(3\Delta^{2}+2\mu^{2}\right)\cos(4\theta_{\textbf{p}})-\cos(2\theta_{\textbf{p}})\left(48\mu^{4}+27p^{4}+8\mu^{2}\left(4\Delta^{2}-9p^{2}\right)\right)\right)\:. \nonumber
\end{align}
The BF surface derived from the above determinant emerges at $\varphi_{\textbf{p}} \in [0 , 2\pi)$ and the following value for $\theta_{\textbf{p}}$
\begin{align}
\theta_{\textbf{p},\rm ferro}
=&
\frac{1}{2}
\arctan \Biggl[\frac{1}{\Delta^{2}p^{4}\left(3\Delta^{2}+2\mu^{2}\right)} \Bigl(27\Delta^{2}p^{6}-72\Delta^{2}\mu^{2}p^{4}+16p^{2}\left(2\Delta^{4}\mu^{2}+3\Delta^{2}\mu^{4}\right) \nonumber \\
&-
\sqrt{2}\Bigl[\Delta^{2}\left(-p^{4}\right)\left(9\mu^{2}p^{2}-4\mu^{4}\right)\bigl(-64\left(16\Delta^{6}+24\Delta^{4}\mu^{2}+12\Delta^{2}\mu^{4}+3\mu^{6}\right)+27p^{6}-12p^{4}\left(12\Delta^{2}+19\mu^{2}\right) \nonumber \\
&+ 48p^{2}\left(8\Delta^{4}+16\Delta^{2}\mu^{2}+11\mu^{4}\right)\bigr)\Bigr]^{1/2}\Bigr),\:
-\frac{1}{\sqrt{\Delta^{2}\left(-p^{4}\right)\left(3\Delta^{2}+2\mu^{2}\right)}}
\Biggl( \Bigl[
\frac{1}{p^{2}\left(3\Delta^{2}+2\mu^{2}\right)}
\big(243p^{10}\left(3\Delta^{2}-2\mu^{2}\right)
\nonumber \\
&-432p^{8}\left(3\Delta^{2}\mu^{2}-10\mu^{4}\right)
-96p^{6}\left(54\Delta^{6}+126\Delta^{4}\mu^{2}+99\Delta^{2}\mu^{4}+118\mu^{6}\right)
-32\sqrt{2}\mu^{2}\left(2\Delta^{2}+3\mu^{2}\right) \times
\\
&
\times \big[\Delta^{2}\left(-p^{4}\right)\left(4\mu^{4}-9\mu^{2}p^{2}\right)\big(64\left(16\Delta^{6}+24\Delta^{4}\mu^{2}+12\Delta^{2}\mu^{4}+3\mu^{6}\right)-27p^{6}+12p^{4}\left(12\Delta^{2}+19\mu^{2}\right) \nonumber
\\
&-48p^{2}\left(8\Delta^{4}+16\Delta^{2}\mu^{2}+11\mu^{4}\right)\big)\big]^{1/2}
+6p^{4}\big(3072\Delta^{6}\mu^{2}+4352\Delta^{4}\mu^{4}+2176\Delta^{2}\mu^{6}+1280\mu^{8} \nonumber \\
&
-9\sqrt{2}\bigl[\Delta^{2}\left(-p^{4}\right)\left(4\mu^{4}-9\mu^{2}p^{2}\right)(64\left(16\Delta^{6}+24\Delta^{4}\mu^{2}+12\Delta^{2}\mu^{4}+3\mu^{6}\right)-27p^{6}+12p^{4}\left(12\Delta^{2}+19\mu^{2}\right)
\nonumber
\\
&
-48p^{2}\left(8\Delta^{4}+16\Delta^{2}\mu^{2}+11\mu^{4}\right))\bigr]^{1/2}\big)-16\mu^{2}p^{2}\bigl(448\Delta^{6}\mu^{2}+576\Delta^{4}\mu^{4}+240\Delta^{2}\mu^{6}+96\mu^{8}
\nonumber
\\
&
-9\sqrt{2}\bigl[\Delta^{2}\left(-p^{4}\right)\left(4\mu^{4}-9\mu^{2}p^{2}\right)\bigl(64\left(16\Delta^{6}+24\Delta^{4}\mu^{2}+12\Delta^{2}\mu^{4}+3\mu^{6}\right)-27p^{6}+12p^{4}\left(12\Delta^{2}+19\mu^{2}\right) \nonumber
\\
&
-48p^{2}\left(8\Delta^{4}+16\Delta^{2}\mu^{2}+11\mu^{4}\right)\bigr)\bigr]^{1/2}\bigr)\big) \Bigl]^{1/2}\Biggr) \Biggr]+\pi c_{1} \:, \nonumber
\end{align}
where $\arctan[x,y]$ denotes the inverse tangent of $y/x$ taking into account which quadrant the point $(x,y)$ is located and $c_1 \in \mathbb{Z}$.
The angle $\theta_{\textbf{p}, \rm ferro}$ is real valued if the absolute value of the momentum $p$ is within the following intervals, for small $\Delta$,
\begin{align}
\frac{2\mu}{3}\le & p\le\frac{2\mu}{3}+\frac{\Delta^{2}}{\mu}\\
2\mu\le & p\le2\mu+\frac{\Delta^{2}}{\mu} \:.
\end{align}
These are then the extents of the two BF surfaces along $(0,0,1)$ direction. As the angle $\theta_{\textbf{p}}$ is increased from zero (or decreased from $\pi$) the two intervals eventually shrink to zero, meanwhile producing a BF surface.

\subsection{Cyclic state}\label{App:cyclis_weak}
The determinant of the cyclic state is given by
\begin{align}
\det(H_{\rm BdG}(\textbf{p}))_{\rm cyclic} & =\mu^{4}\left(2\Delta^{2}+\mu^{2}\right)^{2}+\frac{81p^{8}}{256}+\frac{9}{16}p^{6}\left(\Delta^{2}-5\mu^{2}\right)+\frac{1}{8}p^{4}\left(2\Delta^{4}+29\Delta^{2}\mu^{2}+59\mu^{4}\right)
\nonumber \\
 & -p^{2}\left(2\Delta^{4}\mu^{2}+9\Delta^{2}\mu^{4}+5\mu^{6}\right)
  +\frac{3}{8}\Delta^{2}\mu^{2}p^{4}\Bigl(8\sin^{4}(\theta_{\textbf{p}})\cos(4\varphi_{\textbf{p}})+4\cos(2\theta_{\textbf{p}})+7\cos(4\theta_{\textbf{p}})\Bigr)\:.
\end{align}
In order to find the zeros of the determinant, we first write $\det(H_{\rm BdG}(\textbf{p}))_{\rm cyclic}=0$ as
\begin{equation}
8\sin^{4}(\theta_{\textbf{p}})\cos(4\varphi_{\textbf{p}})+4\cos(2\theta_{\textbf{p}})+7\cos(4\theta_{\textbf{p}}) =-R(p,\mu,\Delta)
\end{equation}
with
\begin{align}
R(p,\mu,\Delta) & =\frac{59\mu^{2}}{3\Delta^{2}}+\frac{2\Delta^{2}}{3\mu^{2}}+\frac{8\mu^{6}}{3\Delta^{2}p^{4}}+\frac{32\Delta^{2}\mu^{2}}{3p^{4}}+\frac{27p^{4}}{32\Delta^{2}\mu^{2}}+\frac{32\mu^{4}}{3p^{4}}
\nonumber \\
 & -\frac{40\mu^{4}}{3\Delta^{2}p^{2}}-\frac{16\Delta^{2}}{3p^{2}}-\frac{15p^{2}}{2\Delta^{2}}-\frac{24\mu^{2}}{p^{2}}+\frac{3p^{2}}{2\mu^{2}}+\frac{29}{3}\:.
\end{align}
We can solve the above equation for $\varphi_{\textbf{p}}$ and find
\begin{equation}
\varphi_{\textbf{p}}=\pm\frac{1}{4}\Biggl[\arccos\Biggl(-\frac{1}{8\sin^4(\theta_{\textbf{p}})}\Bigl[R(p,\mu,\Delta)+4\cos(2\theta_{\textbf{p}})+7\cos(4\theta_{\textbf{p}})\Bigr]\Biggr)+2\pi\Biggr].
\end{equation}
This expression implies restrictions on the parameters $R,p,\theta_{\textbf{p}}$,
since the argument $x$ of the inverse cosine function has to be in
the interval $-1\le x\le1$. This sets the following limits on the
variable $\theta_{\textbf{p}}$:
\begin{align}
\arctan\left(\sqrt{4-\sqrt{31-3R}},\sqrt{8-\sqrt{31-3R}}\right)\le \theta_{\textbf{p},1} & \le\arctan\left(\sqrt{4-\sqrt{31-3R}},\sqrt{\sqrt{31-3R}+8}\right)\\
\arctan\left(-\sqrt{4-\sqrt{31-3R}},\sqrt{\sqrt{31-3R}+8}\right)\le \theta_{\textbf{p},2} & \leq\arctan\left(-\sqrt{4-\sqrt{31-3R}},\sqrt{8-\sqrt{31-R}}\right)\\
\arctan\left(-\sqrt{\sqrt{31-3R}+4},-\sqrt{8-\sqrt{31-3R}}\right)\le \theta_{\textbf{p},3} & \le\arctan\left(-\sqrt{4-\sqrt{31-3R}},-\sqrt{\sqrt{31-3R}+8}\right)\\
\arctan\left(\sqrt{4-\sqrt{31-3R}},-\sqrt{\sqrt{31-3R}+8}\right)\le \theta_{\textbf{p},4} & \le\arctan\left(\sqrt{\sqrt{31-3R}+4},-\sqrt{8-\sqrt{31-3R}}\right)\:.
\end{align}
\end{widetext}
(Here $\phi=\text{arctan}(x,y)$ is the $\text{artcan}$ of $y/x$ such that $x=\cos \phi$ and $y=\sin \phi$.) These equations, on the other hand, yield further constraints since the square root under the
square root has to be real. Hence the function $R(p,\mu,\Delta)$
has to be in the interval
\begin{equation}
5\le R(p,\mu,\Delta)\le\frac{31}{3}.
\end{equation}
Furthermore, for $\mu\gg\Delta,$ also $R>5$. Solving the equation $R=31/3$, which corresponds to the four diagonal directions  $(1,1,1)$,  $(-1,1,1)$, $(1,-1,1)$, and $(1,1,-1)$, and Taylor expanding the result for small $\Delta$, we find
\begin{align}
\frac{2\mu}{3}+\frac{\Delta^{2}}{3\mu}\le p_{1}\le & \frac{2\mu}{3}+\frac{4\Delta^{2}}{3\mu}\\
2\mu-\frac{\Delta^{2}}{\mu}\le p_{2}\le & 2\mu\:.
\end{align}
The nontrivial angular dependence of the BF surfaces that develops as the direction changes from the diagonals in the cyclic state is shown in Fig. \ref{Fig4}.

\end{appendix}

\bibliography{refs_complex}

\end{document}